\documentclass[12pt]{iopart}
\expandafter\let\csname equation*\endcsname\relax
\expandafter\let\csname endequation*\endcsname\relax

\usepackage[utf8]{inputenc}

\usepackage{amssymb}
\usepackage{thmtools}
\usepackage{physics}
\usepackage{graphicx}
\usepackage{subcaption}
\usepackage[figuresleft]{rotating}
\usepackage{placeins}

\usepackage{cite}

\newtheorem{observation}{Observation}

\newcommand{\upd}[1]{^\mathrm{#1}}
\newcommand{\ind}[1]{_\mathrm{#1}}
\newcommand{\sgn}[1]{\mathrm{sgn}{(#1)}}

\begin{document}

\title[Mesoscopic Impurities in GHD]{Mesoscopic Impurities in Generalized Hydrodynamics}
\author{Friedrich Hübner}
\address{Department of Mathematics, King’s College London, Strand, WC2R 2LS London, UK}
\ead{friedrich.huebner@kcl.ac.uk}

\begin{abstract}
We study impurities in integrable models from the viewpoint of generalized hydrodynamics (GHD). An impurity can be thought of as a boundary condition for the GHD equation, relating the state on the left and right side. We find that in interacting models it is not possible to disentangle incoming and outgoing states, which means that it is not possible to think of scattering as a mapping which maps the incoming state to the outgoing state. We then introduce a novel class of impurities, dubbed mesoscopic impurities, whose spatial size is mesoscopic (i.e.\ their size $L_{\mathrm{micro}} \ll L_{\mathrm{imp}} \ll L$ is much larger than the microscopic length scale $L_{\mathrm{micro}}$, but much smaller than the macroscopic scale $L$). Due to their large size it is possible to describe mesoscopic impurities via GHD. This simplification allows one to study these impurities both analytically and numerically. These impurities show interesting non-perturbative scattering behavior, for instance non-uniqueness of solutions and a non-analytic dependence on the impurity strength. In models with one quasi-particle species and a scattering phase shift that depends on the difference of momenta only, we find that one can describe the scattering using an effective Hamiltonian. This Hamiltonian is dressed due to the interaction between particles and satisfies a self consistency fixed point equation. On the example of the hard rods model we demonstrate how this fixed point equation can be used to find almost explicit solutions to the scattering problem by reducing it to a two-dimensional system of equations which can be solved numerically.
\end{abstract}

\maketitle

\section{Introduction}
Integrable models are an important set of fine-tuned (mostly one-dimensional) models in theoretical physics which can be solved exactly. This allows for the analytical study of the effect of interactions in these quantum or classical many-body systems in much greater detail than it would be possible in generic models. At first, it may seem that these fine-tuned models are not necessarily good models for actual physical systems. In particular, they show the unusual feature that they posses infinitely many (local) conserved quantities, while a generic system typically has only 3: particle number, momentum and energy. For instance, this manifests in the equilibrium the system approaches after a long time: it is characterized by a generalized Gibbs ensemble (GGE)~\cite{PhysRevLett.98.050405,PhysRevLett.115.157201}, which takes into account all conserved quantities instead of the usual Gibbs ensembles. This failure of thermalization has been observed in experiments, such as the famous quantum-Newton cradle experiment~\cite{Kinoshita2006} and others~\cite{doi:10.1126/science.1257026,Wang2022}. However, in one dimension, also non-integrable systems often show effects linked to integrability (like long equilibration times, the most famous example here is the FPUT system~\cite{10.1063/1.1855036}). The reason for that is that often these systems are close to an integrable model and therefore for short times they can be treated as integrable models (this also explains why experiments -- which are never precisely fine-tuned -- still can observe GGE's). Only after waiting for long times eventually integrability breaking effects will be observed. Therefore understanding the physics of integrable models provides a first step to understanding the physics of general many-body systems in one dimension. This idea has sparked huge interest in integrable systems, and in particular there has been plenty of work on how to perturb these systems with integrability breaking perturbations, see for instance~\cite{Groha_2017,Collura_2013,Mori_2018,doi:10.1146/annurev-conmatphys-031016-025451,PhysRevLett.115.180601}. 

In the last decade the progress on understanding integrable models was further fueled due to the establishment of a powerful theoretical toolbox: generalized hydrodynamics (GHD)~\cite{PhysRevX.6.041065,PhysRevLett.117.207201}. Generalized hydrodynamics is a framework to study large scale dynamics of integrable models (for a pedagogical introduction, see~\cite{BenGHD}). This provides access to the out-of-equilibrium dynamics of these models, which is particularly useful to compare with experiments~\cite{PhysRevLett.122.090601,10.21468/SciPostPhys.6.6.070,doi:10.1126/science.abf0147,Bouchoule_2022}. GHD is a coarse-grained theory which describes the state of a system locally only by its conversed quantities, all other degrees of freedom are averaged out. This provides an ideal starting ground for studying integrability breaking, since integrability breaking manifests through the breaking of conservation laws. In order to be able to account for an integrability breaking effect in GHD it has to be small in some sense. Small could for instance mean a) that the coupling to the integrability breaking terms is weak. A perturbative treatment typically gives rise to Boltzmann-type equations and has been used to describe various experimentally relevant effects, like the effect of particle losses~\cite{10.21468/SciPostPhys.9.4.044,10.21468/SciPostPhys.12.1.044}, the effect of being only a quasi 1D system~\cite{PhysRevLett.126.090602} and an integrability breaking background~\cite{PhysRevLett.127.130601}. In GHD it is also possible to go beyond weak perturbations and to consider certain strong perturbations, which b) vary slowly in space and time. This includes the experimentally important case of an external trapping potential~\cite{10.21468/SciPostPhys.2.2.014,PhysRevLett.127.130601,Bastianello_2021} or also slowly varying interactions~\cite{PhysRevLett.123.130602}.

Another option to perturb an integrable model is c) to introduce impurities: Impurities can be potentially strong perturbations, but they will only affect the system in a small localized region. Therefore, the system can still be described by GHD away from the impurity, but at the position of the impurity there will be a non-trivial boundary condition relating the state to the left and the right of the impurity. Due to their local nature, it has been postulated and observed that they not lead to thermalization at late times~\cite{Bastianello_2019,Fagotti_2017,Bastianello_2019,10.21468/SciPostPhys.12.2.060}. Impurities have been long studied in the context of integrable models both analytically~\cite{PhysRevLett.121.090404,CASTROALVAREDO2003449,doi:10.1142/S0219887808003223,DELFINO1994518,saleur1998lectures,KONIK1999587,Doikou2013} starting from the infamous Kondo impurity (see for instance the review~\cite{kondo}) and experimentally~\cite{Mebrahtu2012,PhysRevB.75.241407,PhysRevLett.99.066801}.

In this paper we study the out-of-equilibrium dynamics of integrable models in the prescence of an impurity from a generalized hydrodynamics perspective. The situation we have in mind is quite general: Given the initial state $\rho(t=0,x,\lambda)$ at time $t=0$ we want to predict the state $\rho(t,x,\lambda)$ at some later time $t$ taking into account the impurity. First, we formalize and unify ideas already mentioned in the literature~\cite{10.21468/SciPostPhys.12.2.060,PhysRevLett.117.130402,rylands2023transport} to establish a coherent picture for the incorporation of impurities into GHD: The boundary conditions correspond to non-equilibrium stationary states (NESS) of the impurity. In our discussion we will find that the classification of impurities can be quite peculiar: Due to their interaction the reflected particles affect the effective velocities of the incoming particles, implying that it is not possible to think of the outgoing state as a function of the incoming state (unless the model is free)\footnote{Note that if one would apply ordinary scattering theory to the impurity, this problem would not occur. There one can define the S-matrix which maps incoming states to outgoing states. The difference is that ordinary scattering theory corresponds to a infinite time limit $t \gg N$, s.t. eventually all particles are non-interacting. In GHD however, we look at times $t \sim N$, implying that particles are still interacting.}. Because of that it is also not clear whether a boundary condition will always exist or whether it is unique (in fact we will give an example where they it is unique). Another problem for actual computations is that even if the NESS would be known, one would still need to map the states on the left and right onto a GGE, which requires the computation of expectation values of charge densities~\cite{PhysRevLett.117.130402}. Unless the model is non-interacting, this is an incredibly cumbersome task. Note that the problems discussed here are not problems connected with specific impurities, but rather problems connected to the interacting nature of the integrable model.

In this paper we would like to put forward the understanding of the peculiarities of the boundary conditions for impurities in interacting models. We already discussed that the effects mentioned above are not present in free models, which were studied extensively in the non-equilibrium context~\cite{Bastianello_2019,PhysRevLett.120.060602,10.21468/SciPostPhys.12.2.060, refId0,10.21468/SciPostPhys.12.1.011}. We also do not expect them to appear the perturbative treatment applicable to weak impurities~\cite{PhysRevB.98.235128,10.21468/SciPostPhys.12.2.060,PhysRevB.46.15233}, since effects like non-uniqueness are often non-perturbative in nature (and indeed we will observe this at an explicit example). They should appear in integrable impurities, but they are quite fine-tuned and analytical computations on them are involved (also note the recent work on how to incorporate them into GHD~\cite{rylands2023transport}). Instead, for our purposes, we are seeking a broad class of impurities, which a) exists for any integrable model, b) includes strong impurities, c) can be treated analytically and numerically and d) can easily be connected to the GHD language. In the second part of this paper we introduce a class of impurities which satisfies all of these requirements: \emph{mesoscopic} impurities.

Mesoscopic impurities are impurities whose spatial scale is mesoscopic, i.e.\ they are very large (and slowly varying) impurities, but still much smaller than the macroscopic scale. We choose them to be linear combinations of charge densities (note that this is a major difference from the impurity studied in ~\cite{10.21468/SciPostPhys.12.2.060}, which is also mesoscopic in size). Due to their large spatial size one can describe such impurities using the framework of generalized hydrodynamics itself. This drastic simplification will allow us to study them both analytically and numerically in efficient manners. They are a universal description for large scale impurities built from charge-densities and can be used to model non-perturbative impurities in interacting integrable models. Furthermore, the nonequilibrium stationary states are directly available in the language of GHD. Beside this we also expect that it should be possible to implement them in actual experiments, for instance in cold atom setups.

Even more, following basic ideas from generalized hydrodynamics, one can view the regime of mesoscopic impurities also as a zero'th order term of an expansion in $1/L\ind{imp}$, where $L\ind{imp}$ is the spatial size of the impurity. In this sense our results can also be a starting point to systematically gain insights into smaller impurities by taking diffusion (or further higher order corrections) to GHD into account~\cite{10.21468/SciPostPhys.6.4.049,PhysRevLett.121.160603,DeNardis_2023}. 

To keep the discussion simple we restrict ourselves to models in which $\varphi(\lambda,\mu) = \varphi(\lambda-\mu)$ is a function of the difference only. This includes important classical and quantum mechanical systems, like hard rods and the Lieb-Liniger model. In these models we find an intuitive description of scattering at mesoscopic impurities: One can view the interacting system as non-interacting particles evolving in an effective Hamiltonian. The effective Hamiltonian is the single particle bare Hamiltonian plus a correction coming from the interaction of the particles. We show that this effective Hamiltonian satisfies a functional fixed point equation, which provides a convenient starting point for further analysis and explicit solutions. In general we find that the solution the the scattering problem at a mesocopic impurity is invariant under local spatial rescaling of the impurity, identically vanishes for a sufficiently weak impurities and is not necessarily unique for given incoming data. Also scattering at the impurity is purely deterministic, meaning that all particles follow deterministic trajectories.

We further demonstrate how the fixed point equation can be used to obtain almost explicit solutions (in the sense that the problem is reduced to a two coupled equations, which have to be solved numerically) to the scattering problem at the example of hard rods scattering at a potential barrier, particles only approaching from one side. In the case where the length of the rods $d>0$ is positive the solution to impurity problem is unique. However, this is not always the case as we demonstrate for hard rods with negative length $d<0$ (negative length corresponds to a time-delay during scattering). We give an explicit example with two possible stable solutions.

We close the paper with a demonstration of our initial assumption of this paper, namely that replacing the impurity by a boundary condition of the GHD equation indeed gives the correct large scale evolution: For a simple impurity we simulate hard rods scattering at an impurity starting from some large scale initial state at time $t=0$. At a later time $t=T$ we compare the result to the result obtained from simulating the GHD with impurity boundary conditions and show that they coincide. For the simulation of the GHD equation and in particular the boundary condition we use an efficient algorithm outlined in \ref{sec:simulation}.

The paper is structured as follows: In Section \ref{sec:overview} we give an overview over the main results in this paper and in Section \ref{sec:general} we give an introduction to GHD and discuss the relation between the impurity and the boundary condition in general. In Section \ref{sec:mesoscopic} we then introduce mesoscopic impurities and derive the effective Hamiltonian and its fixed point equation in Section \ref{sec:stationary}. In Section \ref{sec:hard_rods} we study the explicit example of hard rods scattering at an potential. In \ref{sec:simulation} we also describe an efficient numerical algorithm to solve the scattering at mesoscopic impurities.

\section{Overview over the main results}
\label{sec:overview}
In this paper we study large-scale dynamics of integrable models in the presence of an mesoscopic impurity constructed from charge-densities of the integrable model $\vu{V} = \int\dd{x} V_n(\tfrac{x}{L\ind{imp}})\vu{q}_n(x)$, where the impurity size $L\ind{imp}$ is mesoscopic, i.e.\ much larger than the microscopic scale, but also much smaller than the macroscopic scale $L$.

In the absence of the impurity the large-scale dynamics of the integrable model are described by the Euler scale generalized hydrodynamics (GHD) equation~\cite{PhysRevX.6.041065,PhysRevLett.117.207201,BenGHD} (see Section \ref{sec:general_GHD}):
\begin{align}
    \partial_t \rho(t,x,\lambda) + \partial_x (v\upd{eff}(t,x,\lambda)\rho(t,x,\lambda)) = 0\label{equ:overview_GHD},
\end{align}
which is interpreted as an evolution equation of quasi-particles with rapidity $\lambda$ and effective velocity $v\upd{eff}$. As we will discuss in Section \ref{sec:general_impurity}, due to the small size of the impurity compared to the macroscopic Euler scale, the impurity will not affect equation \eqref{equ:overview_GHD} anywhere except at the impurity location $x=0$. This can be described by adding a boundary conditions to the GHD equation at $x=0$, which relate the state to the left and the right of the impurity. This boundary condition depends on the specific impurity and is connected to its non-equilibrium stationary states. Constructing these non-equilibrium stationary states for a general impurity is a hard problem.

\subsection{Mesoscopic impurities}

The main part of this paper studies the relation between the impurity and its corresponding boundary condition for the mesoscopic impurities: Due to their large size compared to the microscopic length scale, one can approximate the dynamics on the impurity by GHD. Under this approximation the problem reduces to finding stationary solutions to the GHD equation with external potential:
\begin{align}
    \partial_x ((\partial_\lambda E)\upd{dr} n) - \partial_\lambda ((\partial_x E)\upd{dr} n) &= 0,
\end{align}
where $E(x,\lambda)$ is the position and rapidity dependent bare quasi-particle energy and $n(x,\lambda)$ is the occupation function. Each solution to this equation (which is on mesoscopic spatial scales $x$), evaluated at $x \to \pm \infty$ corresponds to a possible boundary condition for the macroscopic GHD equation \eqref{equ:overview_GHD}.

In models where the scattering phase shift $\varphi(\lambda-\mu) = \varphi(\lambda-\mu)$ is a function of the rapidity difference only, we show that the trajectories of the interacting particles at the impurity can be thought of as trajectories of non-interacting particles evolving according to an effective Hamiltonian $H(x,\lambda)$, see Section \ref{sec:stationary}, i.e.\ particles follow trajectories given by:
\begin{align}
    \dv{x(t)}{t} &= \partial_\lambda H(x(t),\lambda(t)) & \dv{\lambda(t)}{t} &= -\partial_x H(x(t),\lambda(t)).
\end{align}
This effective Hamiltonian differs from the bare single particle Hamiltonian $E(x,\lambda)$ due to a contribution from scattering between particles. In the simplest case the relation reads:
\begin{align}
    H(x,\lambda) = E(x,\lambda) + \vu{T}N(x,\lambda)\label{equ:motivation_H_N},
\end{align}
where $\vu{T}$ is given by the scattering phase shifts of particles $T(\lambda-\mu) = \frac{\varphi(\lambda-\mu)}{2\pi}$ and $N(x,\lambda)$ is defined by $\grad N(x,\lambda) = n(x,\lambda) \grad H$. The occupation function $n(x,\lambda)$ is transported along the trajectories of particles (the same is true for $N(x,\lambda)$). Thus, given $H(x,\lambda)$ we can construct $N(x,\lambda)$ by transporting the incoming data from the left and right side along the particle trajectories. On the other hand, given $N(x,\lambda)$ one can compute $H(x,\lambda)$ via \eqref{equ:motivation_H_N}. Therefore $H(x,\lambda)$ and $N(x,\lambda)$ have to be determined self-consistently. We use this to construct iterative schemes to solve the scattering problem numerically (see \ref{sec:simulation}). On the analytical side we show how to simplify the self-consistency relation into a functional fixed point problem for $H(x,\lambda)$ only (see Section \ref{sec:stationary_scattering_problem}).

In general we establish the following properties of mesoscopic impurities in general integrable model:
\begin{itemize}
    \item Scattering at mesoscopic impurities is invariant under local spatial rescalings of the impurity.
    \item Scattering identically vanishes if the strength of the impurity is below a certain cutoff (meaning that scattering is non-perturbative in impurity strength).
    \item The trajectory of particles at the impurity are deterministic. This means that all particles with a certain rapidity $\lambda$ are either transmitted or reflected.
\end{itemize}

\subsection{Hard rods with rapidity independent potential}
In one part of this paper we demonstrate how one can use this functional fixed point equation to find analytical solutions to the scattering problem in the hard rods model (rod length $d$) with a simple impurity $V(x)$, which does not depend on rapidity (see Section \ref{sec:hard_rods}), with incoming particles only from the left. The incoming particle distribution is characterized by an arbitrary occupation function $n\ind{L}(\lambda)$ on the left side. In this case the effective Hamiltonian has a particularly simple form:
\begin{align}
    H(x,\lambda) &= \frac{1}{2}(\lambda-d \expval{\lambda})^2 + W(x)\label{equ:overview_hardrods_H}.
\end{align}
Here $d \expval{\lambda} = d\int\dd{\lambda} \lambda \rho(\lambda)$ is a (position independent) momentum shift and $W(x)$ is an effective potential defined via 
\begin{align}
    \partial_x W(x) = 1\upd{dr}(x)\partial_x V(x) = \frac{\partial_x V}{1+d\int\dd{\lambda}n(x,\lambda)}.
\end{align}
Note that this effective potential $W(x)$ is lower than the bare potential $V(x)$ due to the interaction between particles in the usual hard rod case $d>0$. This means that the presence of other particles help other particles to overcome the barrier. Intuitively speaking this is because an incoming tracer particles will occasionally jump over other particles, thereby skipping some parts of the uphill slope. We also discuss the case $d<0$ where the effective potential is increased compared to the bare potential. Up to the momentum shift, \eqref{equ:overview_hardrods_H} is just the Hamiltonian of a single particle scattering at some potential barrier, whose scattering problem can be solved easily: A particle is transmitted if its kinetic energy is above the maximum value $\bar{W}$ of $W(x)$, otherwise reflected. For given $d\expval{\lambda}$ and $\bar{W}$ this determines a distribution of particles $n(x,\lambda)$ at the impurity. From this distribution one can compute $d\expval{\lambda}$ and $\bar{W}$ using the formulas above. This gives rise to the following set of self-consistency equations (see Section \ref{sec:hard_rods_example}):
\begin{align}
    \bar{V} &= \bar{W} + \tfrac{d}{2\pi} \Big[\int_{d\expval{\lambda}}^\infty\dd{\lambda} N\upd{L}(\tfrac{1}{2}(\lambda-d\expval{\lambda})^2 + \bar{W})\nonumber\\
    &\indent- \int_{d\expval{\lambda}-\sqrt{2\bar{W}}}^\infty\dd{\lambda} N\upd{L}(\tfrac{1}{2}(\lambda-d\expval{\lambda})^2)+ \sqrt{2\bar{W}}N\upd{L}(\bar{W})\Big]\\
    d\expval{\lambda} &= d\qty[N\upd{L}(\infty)-N\upd{L}(\bar{W})].
\end{align}
Here $N\upd{L}(h)$ depends on the incoming particle occupation function $n\ind{L}(\lambda)$ as follows:
\begin{align}
    N\upd{L}(h) &= \int_{0}^{\sqrt{2h}}\dd{v} v n\ind{L}(d\expval{\lambda}+v).
\end{align}

\subsection{Non-uniqueness of the boundary conditions}
We use this analytical solutions in the hard rods model to explicitly demonstrate the following intriguing feature of scattering at impurities in integrable models:
\begin{itemize}
    \item The solution to the scattering problem (i.e.\ the boundary condition) is not necessarily unique for given prescribed incoming data.
\end{itemize}
In fact for the hard rods model with $d>0$ the solutions is always unique for our specific impurity, however for $d<0$ we discuss an example where two stable (and one unstable) solution exist (see Section \ref{sec:hard_rods_hysteris}).

The last point makes it clear that the boundary condition describing the impurity in the GHD formalism is fairly different from another formalism to study impurities: The S matrix. The S matrix is defined as a map that maps incoming states to their corresponding outgoing states. This is always well-defined and unqiue. The difference is a matter of time scales: The S matrix formalism describes the scattering using asymptotic states at very long time times, where all particles are sufficiently separated such that they can be treated as non-interacting particles. The boundary condition in the GHD formalism on the other hand describes scattering at Euler scale times, where particles are still interacting.

We would like to stress that this non-uniqueness is not a feature of a specific impurity, but is due to the interaction in the integrable model: The fundamental reason is that in interacting models the incoming particles interact with the outgoing particles. In fact in general it is not even possible to specify the incoming state a priori, since this might be altered by outgoing state (for our discussion of the hard rods we use incoming data where the incoming state is unambiguous), see discussion in Section \ref{sec:general_impurity}.

\subsection{Simulating the GHD equation with boundary conditions}
\begin{figure}[!h]
    \centering
    \includegraphics[width=\textwidth]{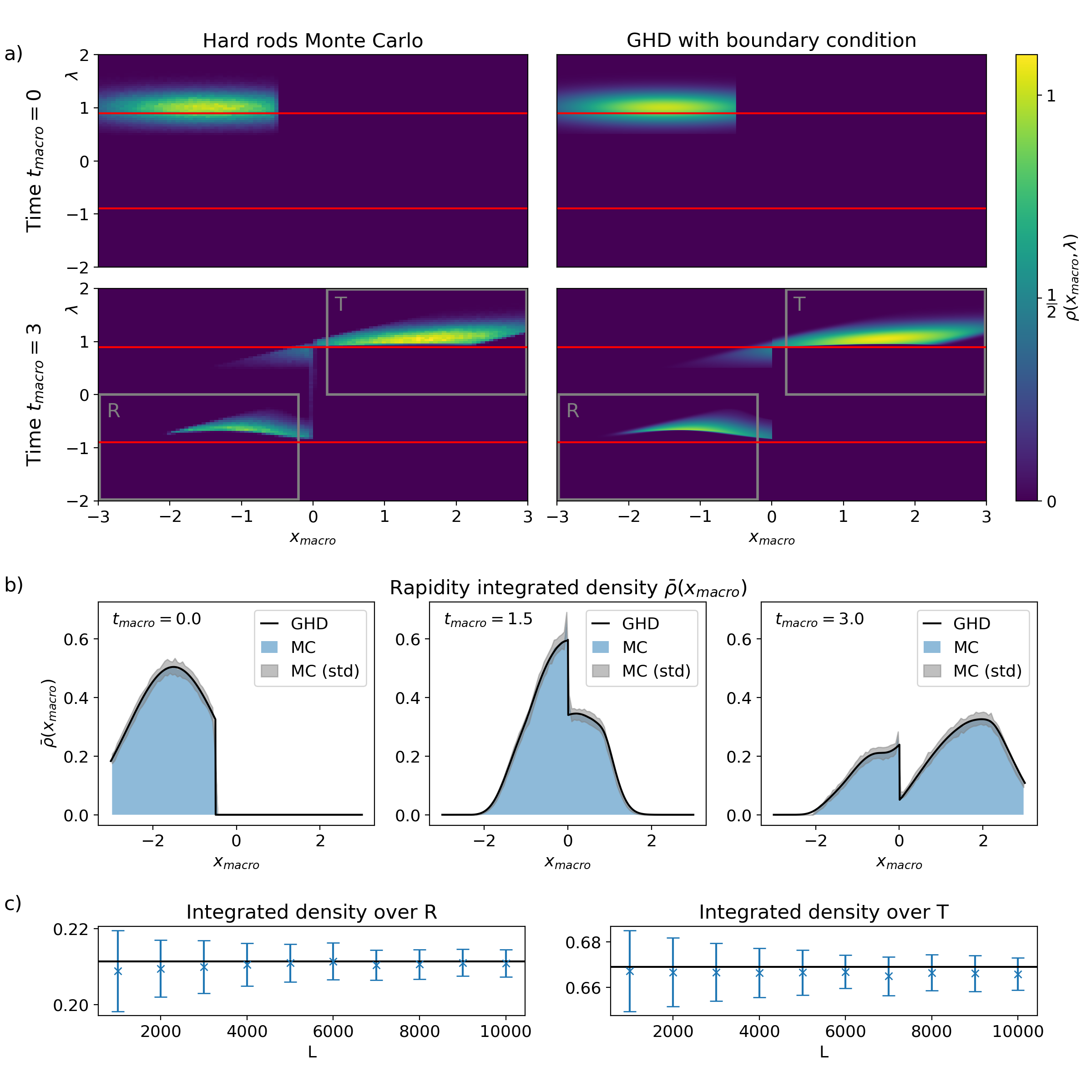}
    \caption{Comparison of microscopic hard rods simulation with $d = 0.3$ (averaged over 100 samples) and the corresponding result obtained by simulation of the GHD equation with impurity boundary conditions starting from the initial state \eqref{equ:hard_rods_comparision_init_state}: a) Comparison of the GHD quasi-particle density $\rho$ with a 2D histogram of the hard rods distribution ($100$ bins in both dimensions). The red lines indicate the momenta $\lambda = \pm\sqrt{2\alpha}$ whose kinetic energy corresponds to the height of the potential $\alpha = 0.4$ (given for comparison with non-interacting particles $d=0$ (as in \ref{sec:simulation_simulation})). b) The rapidity integrated density at three different times -- comparison of the GHD evolution (black line) with the hard rods simulations (blue, again $100$ bins in $x$ direction) and corresponding standard deviation (gray). c) Convergence as $L \to \infty$ of the quasi-particle density $\rho$ integrated over the regions R (reflected particles) and T (transmitted particles) indicated in a) (solid line -- GHD result, data points -- mean and standard deviation of hard rods simulations).}
    \label{fig:mc_GHD_comparision}
\end{figure}

We further demonstrate numerically for the hard rods model that the proposed formalism (GHD equation + boundary condition) indeed describes the out-of-equilibrium dynamics of integrable models in the Euler scaling limit. We simulate the scattering of hard rods at a mesoscopic impurity $V(x\ind{micro}) = \alpha \exp(-\tfrac{1}{2}(x\ind{micro}/L\ind{imp})^2)$ with $L\ind{imp} = \tfrac{1}{10}L^{\tfrac{3}{4}}$ for a very large $L=10000$ starting from an initial state given by \eqref{equ:hard_rods_comparision_init_state}. This is done two ways: First, we simulate the microscopic hard rod dynamics starting from 100 randomly generated initial states. Second, we simulate the GHD equation together with the impurity boundary condition. In Figure \ref{fig:mc_GHD_comparision} we compare the resulting distribution of particles at macroscopic time $t\ind{macro} = 3$ of both simulations and show that they agree. For details on the numerical simulations see Section \ref{sec:comparison}.
\FloatBarrier

\section{General considerations}
\label{sec:general}
Given a system which consists of an (translation invariant) integrable model and an impurity we would like to understand the behavior of the system on large scales. First, let us specify what we mean by an impurity: An impurity is a (possibly strong) localized perturbation of the integrable model whose spatial size $L\ind{imp}$ is much smaller than the macroscopic scale $L \gg L\ind{imp}$. This is contrary to the case of an external potential, which is a perturbation whose spatial size $L\ind{ext}$ is comparable to the macroscopic scale $L \sim L\ind{ext}$. In this paper we will restrict ourselves to the case where the system far on the left and far on the right is equivalent. However, the same discussion and the tools can be extended to the case where the system on both sides is different. A simple example of that is the case of a potential $V(x)$ which decays to different values $V(\infty) \neq V(-\infty)$ as $x \to \pm \infty$. A more complicated example would be systems with different interactions on both sides of the impurity. In such a situation the impurity acts as the connection between two different systems. Also note that our analysis also applies to boundaries (seen as a limiting case where one system is non-existent).

\subsection{Generalized Hydrodynamics}
\label{sec:general_GHD}
In this paper we want to study the effect of an impurity in a non-equilibrium setting. The problem is as follows: Given an initial state of the system at time $t=0$, predict the state of the system at some later time. In interacting integrable models already the evolution without any impurity is complicated and in most cases cannot be computed explicitly. However, in many interesting cases and experiments (for instance in cold atom systems~\cite{PhysRevLett.122.090601,10.21468/SciPostPhys.6.6.070,doi:10.1126/science.abf0147}) the macroscopic scale $L$ is much larger than the microscopic length scale of the particle, the system is observed on long timescales compared to the microscopic timescale and also the system contains many particles. In this case one can describe the out-of-equilibrium dynamics on large scales integrable models by generalized hydrodynamics~\cite{PhysRevX.6.041065,PhysRevLett.117.207201,BenGHD}:

The starting point for the derivation of GHD are the microscopic conservation laws:
\begin{align}
    \partial_t \expval{\vu{q}_n(t,x)} + \partial_x\expval{\vu{j}_n(t,x)}.\label{equ:general_microscopic_conservation_law}
\end{align}
Here $\expval{\vu{q}_n(t,x)}$ are the expectation values of the charge-densities of the conserved charges $\vu{Q}_n = \int\dd{x}\vu{q}_n(x)$ (labelled by some indices $n$) and $\expval{\vu{j}_n(t,x)}$ are the expectation values of the associated currents. Consider now a situation in which the system is observed on a length scale (the `macroscopic length scale' $L$) much larger than the microscopic length scale $L \gg L\ind{micro} \sim \order{1}$ and also on a time scale (the `macroscopic time scale' $T$) much larger than the microscopic time scale $T \gg T\ind{micro} \sim \order{1}$. This allows to average $\expval{\vu{q}_n(t,x)}$ and $\expval{\vu{j}_n(t,x)}$ over a mesoscopic fluid cell in space and time:
\begin{align}
    q_n(t,x) &= \frac{1}{L\ind{meso}T\ind{meso}}\int_{Lx-L\ind{meso}/2}^{Lx+L\ind{meso}/2}\dd{y}\int_{Tt-T\ind{meso}/2}^{Tt+T\ind{meso}/2}\dd{s} \expval{\vu{q}_n(s,y)}\label{equ:general_fluidcell_q}\\
    j_n(t,x) &= \frac{1}{L\ind{meso}T\ind{meso}}\int_{Lx-L\ind{meso}/2}^{Lx+L\ind{meso}/2}\dd{y}\int_{Tt-T\ind{meso}/2}^{Tt+T\ind{meso}/2}\dd{s} \expval{\vu{j}_n(s,y)}\label{equ:general_fluidcell_j}.
\end{align}
Note that $q_n(t,x)$ and $j_n(t,x)$ are now expressed in terms of macroscopic coordinates $t$ and $x$. Here $L\ind{micro} \ll L\ind{meso} \ll L$ and $T\ind{micro} \ll T\ind{meso} \ll T$ are intermediate mesoscopic scales. Since the mesoscopic scales are large compared to the microscopic scale, we expect equilibration of the state to a generalized Gibbs ensemble (GGE) to happen inside the Fluid cell and \eqref{equ:general_fluidcell_q} and \eqref{equ:general_fluidcell_j} to be given by the corresponding expectation values of this GGE. Since a GGE is fully characterized by its expectation values on all the charges, we can in principle express the currents $j_n(t,x) = j_n[\qty{q_m(t,x)}]$ as functions of the charge densities. This finally yields the following:
\begin{align}
    \partial_t q_n(t,x) + \partial_x j_n[\qty{q_m(t,x)}] = 0.\label{equ:general_GHD_q_n}
\end{align}
This is a closed set of equations describing the time evolution of the charge densities. Mathematically speaking equation \eqref{equ:general_GHD_q_n} is expected to hold in the Euler scaling limit $L \sim T \sim N \to \infty$. Here $N$ is the total particle number, which is scaled linearly with the macroscopic length scale in order to ensure that states can locally be described by finite density GGE. For further details on the derivation and the Euler scaling limit, see the original papers~\cite{PhysRevX.6.041065,PhysRevLett.117.207201} and the lecture notes~\cite{BenGHD}.

In the context of this paper we will work in the Euler scaling limit. We want to stress that the our results and discussions are only valid in this limit: They will not describe the system at short or long times compared to the Euler scale time. Also we will not consider any corrections in $1/L$ (like diffusive corrections~\cite{10.21468/SciPostPhys.6.4.049,PhysRevLett.121.160603,DeNardis_2023}).

In order to apply \eqref{equ:general_GHD_q_n} it is required to know the currents $j_n[\qty{q_m}]$ as functions of any charge density. This relation can be stated in a convenient form by choosing a specific `basis' on the space of all conserved quantities $q_n(t,x) = \int\dd{\lambda}\rho(t,x,\lambda)h_n(\lambda)$, where $h_n(\lambda)$ is interpreted as the $n$'th bare charge carried by a single quasi-particle with rapidity $\lambda$ (these quasi-particles correspond to asymptotic states of the integrable model~\cite{PhysRevX.6.041065,PhysRevLett.117.207201,BenGHD}). In many Galilean invariant models (such as the hard rods model and the Lieb-Liniger model) the rapidity equals the momentum $P(\lambda)=\lambda$,  typically the first three of these charges are given by $h_0(\lambda)=1, h_1(\lambda)=\lambda$ and $h_2(\lambda)=\tfrac{\lambda^2}{2}$, which represent the particle number, momentum and energy of a quasi-particle. More generally the (bare) momentum of a particle is a function $P(\lambda)$ of the rapidity. Similarly, each quasi-particle also has a (bare) energy $E(\lambda)$. There also exist models with multiple species of quasi-particles. For convenience, in this work we will restrict ourselves to the simplest case of a single particle species. However, the results can easily be adapted.

In the basis $h_n(\lambda)$ the (Euler scale) generalized hydrodynamics equation becomes~\cite{PhysRevX.6.041065,PhysRevLett.117.207201,BenGHD}
\begin{align}
    \partial_t \rho(t,x,\lambda) + \partial_x (v\upd{eff}(t,x,\lambda)\rho(t,x,\lambda)) &= 0\label{equ:general_GHD_conservation},  
\end{align}
where $v\upd{eff}(t,x,\lambda) = \frac{j(t,x,\lambda)}{\rho(t,x,\lambda)}$ is interpreted as effective velocity of the particles and satisfies the following equation:
\begin{align}
    v\upd{eff}(t,x,\lambda) = \frac{E'(\lambda)}{P'(\lambda)} + \int\dd{\mu}\frac{\varphi(\lambda,\mu)}{P'(\lambda)}\rho(t,x,\mu)(v\upd{eff}(t,x,\mu)-v\upd{eff}(t,x,\lambda)).\label{equ:general_GHD_effective_velocity}
\end{align}
This self-consistency equation has an intuitive interpretation: The effective velocity is given by the bare velocity $v(\lambda)=\dv{E}{p} = \frac{E'(\lambda)}{P'(\lambda)}$ ($'$ denotes the derivative w.r.t. $\lambda$) plus a contribution from the scattering of the quasi-particles. During scattering of two quasi-particles with rapidity $\lambda$ and $\mu$ their trajectories are shifted by the scattering phase shift $\varphi(\lambda,\mu)$ (which we will assume to be symmetric). When $\varphi(\lambda,\mu)$ is positive the interaction corresponds to a time-delay or a backwards displacement of the trajectory of the particle. When $\varphi(\lambda,\mu)$ is negative the interaction speeds up the particle and thus corresponds to a forward displacement of the trajectories. These kind of particle dynamics are called tracer dynamics and the corresponding quasi-particles are referred to as tracer particles (for a discussion see~\cite{BenGHD}, Section 3.1). 

This tracer dynamics gives a unified picture for the large scale dynamics of integrable models, however, their microscopic origin differs from model to model. The simplest model to understand them is the hard rods model (which we will also use as an example later in the paper): 
This model describes hard spheres of diameter $d$ in one-dimension. Its $N$ particle Hamiltonian is given by:
\begin{align}
    H = \sum_{i=1}^N \tfrac{\lambda_i^2}{2} + \sum_{i,j=1}^N V(x_i-x_j),
\end{align}
with interaction potential $V(x) = \infty$ if $|x|<d$ and $V(x)=0$ otherwise. During scattering the physical particles instantaneously swap their momenta. The corresponding tracer particles keep their momentum during collision, but swap their positions instead (this is merely a relabeling of particles). Therefore a tracer particle will retain its momentum for all times. In this model $P(\lambda) = \lambda$, $E(\lambda) = \tfrac{1}{2}\lambda^2$ and the scattering phase shift is simply $\varphi(\lambda,\mu)=-d$. Similar ideas can be used to define the tracer dynamics for other classical models (see~\cite{BenGHD}, Section 3.1). In the context of integrable PDE's instead the tracer dynamics are given by solitons (a wave that preserves its shape during evolution), which also obtain an effective position shift during scattering~\cite{el2003thermodynamic,el2021soliton}. In quantum mechanical models the correspondence of tracer dynamics to the microscopic dynamics is not well understood in general (see recent progress~\cite{doyon2023ab}).

Equation \eqref{equ:general_GHD_conservation} is the GHD equation in conservation form. There also exists a formulation of the GHD equation as a transport equation (this form will be most relevant in the context of impurities as we will discuss in the next section). Define the so called occupation function:
\begin{align}
    n(t,x,\lambda) &= 2\pi\frac{\rho(t,x,\lambda)}{P'(\lambda)+\int\dd{\mu}\varphi(\lambda,\mu)\rho(t,x,\mu)},
\end{align}
which can be shown to satisfy the following transport equation~\cite{PhysRevX.6.041065}:
\begin{align}
    \partial_t n(t,x,\lambda) + v\upd{eff}(t,x,\lambda)\partial_x n(t,x,\lambda) &= 0.
\end{align}
In this formulation $n(t,x,\lambda)$ is transported along the trajectories of the quasi-particles (called GHD characteristics) with velocity $v\upd{eff}$. One can express $\rho$ and $v\upd{eff}$ in terms of the occupation function as
\begin{align}
    \rho &= \frac{1}{2\pi}{P'}\upd{dr} n & v\upd{eff} &= \frac{{E'}\upd{dr}}{{P'}\upd{dr}},
\end{align}
where the dressing operation applied on a function $f(\lambda)$ is defined as the solution to:
\begin{align}
    f\upd{dr}(\lambda) = f(\lambda) + \int\tfrac{\dd{\mu}}{2\pi} \varphi(\lambda,\mu) n(t,x,\mu) f\upd{dr}(\mu) = f(\lambda) + \vu{T} n(t,x,\mu) f\upd{dr}(\mu)\label{equ:general_dressing}.
\end{align}
Here we defined the operator $\vu{T}$, whose kernel $T(\lambda,\mu) = \tfrac{1}{2\pi}\varphi(\lambda,\mu)$ is given by the scattering phase shift, which depends on the model. Note that the above equations are written for an integrable model with a single particle species. It is also possible to formulate them for multiple particle species~\cite{PhysRevX.6.041065,PhysRevLett.117.207201}.

GHD can also be extended to situations where the charge-densities are coupled to slowly varying external potentials, i.e. $\vu{V} = \int\dd{x} V_n(\tfrac{x}{L})\vu{q}_n(x)$~\cite{10.21468/SciPostPhys.2.2.014}:
\begin{align}
	\partial_t \rho(t,x,\lambda) + \partial_x\qty(\frac{(\partial_\lambda E)\upd{dr}(t,x,\lambda)}{{P'}\upd{dr}(t,x,\lambda)}\rho(t,x,\lambda)) -\partial_\lambda\qty(\frac{(\partial_x E)\upd{dr}(t,x,\lambda)}{{P'}\upd{dr}(t,x,\lambda)} \rho(t,x,\lambda)) = 0.\label{equ:general_GHD_external_potential}
\end{align}
Here $E(x,\lambda) = E_0(\lambda) + \sum_n V_n(x)h_n(\lambda)$ is the position dependent bare energy of a quasi-particle with rapidity $\lambda$ (here $E_0(\lambda)$ denotes the single particle kinetic energy). In case the phase shift $\varphi(\lambda,\mu) = \varphi(\lambda-\mu)$ is a function of the difference $\lambda-\mu$ only, this can also be written in transport form:
\begin{align}
    \partial_t n(t,x,\lambda) + \frac{(\partial_\lambda E)\upd{dr}(t,x,\lambda)}{{P'}\upd{dr}(t,x,\lambda)}\partial_xn(t,x,\lambda) -\frac{(\partial_x E)\upd{dr}(t,x,\lambda)}{{P'}\upd{dr}(t,x,\lambda)}\partial_\lambda n(t,x,\lambda) = 0.
\end{align}

\subsection{The effect of an impurity in GHD}
\label{sec:general_impurity}
So far we discussed the system without an impurity. By definition the impurity is way smaller than the macroscopic scale and thus from the GHD viewpoint the impurity will shrink to a point. Throughout this paper we will place the impurity at $x = 0$. For $x \neq 0$ the system is still described by the GHD equation, but at $x = 0$ the impurity will lead to some boundary condition, which relates $n\ind{L}(\lambda) = n(0^-,\lambda)$ and $n\ind{R}(\lambda) = n(0^+,\lambda)$. We can now write the GHD equation (e.g.\ in transport form) including the impurity as:
\begin{align}
    \begin{cases}\partial_t n + v\upd{eff} \partial_x n = 0 & x\neq 0\\
    (n\ind{L},n\ind{R}) \in \mathcal{M} & x = 0,
    \end{cases}
\end{align}
where $\mathcal{M}$ denotes the set of all physically allowed relations between $n\ind{L}$ and $n\ind{R}$. Note that we choose to express the boundary conditions in terms of the occupation function $n$ and not $\rho$ (see discussion at the end of this section).

How can one characterize the set $\mathcal{M}$? For that we need to study the scattering problem at the impurity. The idea is that since the impurity is small $L\ind{imp} \ll L$, the timescales for scattering at the impurity are also way smaller than the macroscopic timescales. Thus, at the scale of the impurity we study a long time limit where we can send $t\ind{imp} \to \infty$. Also since $n\ind{L}$ and $n\ind{R}$ change on the macroscopic time-scale (i.e.\ slowly), we can assume them as constant during scattering. This is the usual assumption on scattering at impurities in general and is, for instance, at the base of scattering theory in quantum mechanics. In the limit $t\ind{imp} \to \infty$ the solution converges to a stationary solution of the equation of motion. A prominent example of this is the Lippmann-Schwinger equation in quantum mechanics where the scattering state is an eigenstate of the Hamiltonian (and thus stationary in time)~\cite{PhysRev.79.469}. We will also see this feature when we study mesoscopic impurities. 

The scattering problem of the impurity therefore consists of finding a solution to the stationary equations of motion of the impurity system, so called non-equilibrium stationary states (NESS). This is well known, also in the context of GHD, and these NESS have been constructed and studied in special cases~\cite{PhysRevLett.117.130402,PhysRevLett.120.060602,10.21468/SciPostPhys.12.2.060}.

Given such a NESS, one can construct the boundary conditions in the following way: Evaluating a NESS far away from the impurity $x\ind{imp} \to \pm \infty$ it will become a solution to the unperturbed system. From the GHD viewpoint this state should correspond to a GGE, which is characterised by its expectation values of the charge densities. Therefore, one has to compute the expectation values of charge densities far away from the impurity. Given those expectation values one can in principle construct a GGE, which corresponds to a quasi-particle density $\rho(\lambda)$ or alternatively to an occupation function $n(\lambda)$ in GHD. The $n(\lambda)$ obtained by this procedure are the boundary conditions $n\ind{L}$ and $n\ind{R}$. To summarize we look for a solution to the stationary impurity system, a NESS whose asymptotics for $x\ind{imp} \to \pm \infty$ are described by $n\ind{R/L}$. The set of all those possible states then determines $\mathcal{M}$.

Usually, one does not describe the boundary condition just as a set $\mathcal{M}$, but one views the outgoing state as a function of the incoming state $n\ind{out}[n\ind{in}]$. For instance, in quantum mechanics one defines the $S$ matrix which maps an incoming state to the corresponding outgoing state. This is a natural and physically intuitive description of the scattering at the impurity. 

Unfortunately, viewing the outgoing state as a function of the incoming state is not possible in interacting models. The fundamental reason is that the effective velocity of particles gets affected by the presence of all other particles, incoming as well as outgoing. Therefore, given for instance the solution $n\ind{L}$ we can determine which particles are incoming and which are outgoing on the left side. But if we do not know the solution we do not know which particles are incoming a priori. In addition to that, even in cases where it is clear which particles are incoming, it is not clear whether the solution to the scattering problem is unique. In fact, we will demonstrate the non-uniqueness at an explicit example, see Section \ref{sec:hard_rods_hysteris}. To avoid potential confusion, this non-uniqueness refers to the non-uniqueness of boundary conditions. If one applies ordinary scattering theory to any impurity, it is always possible to define a S-matrix, which maps incoming states to outgoing states in a unique way. However, ordinary scattering theory is based on a long time limit and describes the incoming and outgoing states in terms of asymptotic states. These asymptotic states form only at very long times $|t| \gg L,N$ where the particles have separated so much that they do not interact anymore. In GHD on the other hand, we consider much shorter times $|t| \sim L,N$: The states to the right and to the left of the impurity are finite density states, where particles are still interacting. In fact, this also shows that the S matrix is not the correct object to describe scattering on hydrodynamic timescales of interacting integrable models.

Due to the peculiarities outlined in the last paragraph, in general, we need to describe the boundary conditions by a set $\mathcal{M}$. However, there are some restrictions on that set based on basic symmetries and conservation laws. For instance, if the impurity is PT symmetric then if $(n\ind{L}(\lambda),n\ind{R}(\lambda)) \in \mathcal{M}$, also $(n\ind{R}(-\lambda),n\ind{L}(-\lambda)) \in \mathcal{M}$. Unless the impurity is integrable as well, the impurity will break some, if not all, conservation laws. For a typical impurity we can expect that it will conserve the total particle number and the total energy. In this case we have the restrictions $\int\dd{\lambda}v\upd{eff}\ind{L}\rho\ind{L}(\lambda) = \int\dd{\lambda}v\upd{eff}\ind{R}\rho\ind{R}(\lambda)$ and $\int\dd{\lambda}E(\lambda)v\upd{eff}\ind{L}(\lambda)\rho\ind{L}(\lambda) = \int\dd{\lambda}E(\lambda)v\upd{eff}\ind{R}(\lambda)\rho\ind{R}(\lambda)$. However, in principle also particle number or energy conservation violating impurities are possible.

We finish this general discussion with another subtlety which only appears in interacting models: In principle one can express the set $\mathcal{M}$ either in terms of $\rho$ or in terms of $n$. Both are possible, however, we find that $n$ is better suited for the following reason: Consider the unperturbed model and take as initial state two bumps: a bump on the left side with positive velocity and a bump on the right side with negative velocity. When the system evolves the two bumps will collide. During the collision $n(\lambda)$ will be transported along some complicated trajectory of the quasi-particle, but the value $n(\lambda)$ will not change. The quasi-particle density $\rho = \tfrac{1}{2\pi}{P'}\upd{dr} n$, however, will change along the trajectory, since it is multiplied by ${P'}\upd{dr}$, which depends on the presence of other particles as well. The same effect appears during the scattering at the impurity: The reflected particles will alter the shape of the incoming $\rho$, but not of the incoming $n$. Therefore $n$ is better suited to describe an incoming state.

\section{Mesoscopic impurites}
\label{sec:mesoscopic}
In general, microscopic impurities can become very complicated and break many conservation laws, which is why we cannot hope to find a general solution to their scattering problem. Usually one can only study very specific integrable impurities or try to study the problem perturbatively in the impurity strength. Even then, in order to connect with GHD, one has to compute the expectation values of all charge densities, which is incredibly cumbersome in interacting models~\cite{PhysRevLett.117.130402}. Therefore, most computations so far restricted to free models. As we discussed, in free models one does not encounter the delicate problems for interacting models outlined in the last section. In order to understand them it would be instructive to have a class of impurities, which can be treated analytically for interacting models and in particular whose solutions can be easily connected to the GHD language. In the remainder of the paper we will introduce such a class of impurities, which we call mesoscopic impurities. This class of impurities exists for any integrable model, therefore allowing to model and study impurity effects quite generally.\\ 

The idea is to study impurities which are large in size (compared to microscopic), but smaller than the macroscopic length scale $L$ of the system, i.e.\ $L\ind{imp} \sim L^\gamma$, where $\tfrac{1}{2} < \gamma < 1$ (we require $\gamma > \tfrac{1}{2}$ in order to avoid diffusive effects). To be precise we consider impurities which are combinations of charge densities $\vu{V} =  \int\dd{x}V_n\qty(\tfrac{x}{L\ind{imp}}) \vu{q}_n(x)$, where $\vu{q}_n(x)$ are the charge densities of the integrable model.

We dub these impurities mesoscopic impurities, since their size is mesoscopic. On the length scale $L\ind{imp}$ we can still divide space in small fluid cells in which we can assume that local relaxation to a GGE has already happened. Since this assumption is the basic assumption of GHD, this in fact means we can fully describe the impurity using the GHD framework alone. This is a great simplification. Of course the results only hold for very large impurities. But still they provide a first insight into general features of impurities in GHD and also can be seen as a first approximation to an actual impurity. In order to study smaller impurities the next step would be to also take diffusive corrections into account, which would give access to effects of impurities on scales $L\ind{imp} \sim L^\gamma$ for $\tfrac{1}{3} < \gamma < \tfrac{1}{2}$. By adding more and more orders of the gradient expansion one can thus access smaller and smaller impurities. However, it is not clear whether the gradient expansion will give rise to a convergent series, so potentially it will fail to describe microscopic impurities.

In order to find an equation describing the mesoscopic impurity consider the general GHD equation for a space dependent bare energy function $E(x,\lambda)$~\cite{10.21468/SciPostPhys.2.2.014}:
\begin{align}
	\partial_t ({P'}\upd{dr}n) + \partial_x((\partial_\lambda E)\upd{dr}n) -\partial_\lambda((\partial_x E)\upd{dr} n) = 0,\label{equ:mesoscopic_derivation_GHD}
\end{align}
which is written in conservation form, but using the occupation function $n$ (compare to \eqref{equ:general_GHD_external_potential}). In our case $E(x,\lambda) = E_0(\lambda) + V(x/\ell,\lambda)$, which is connected to $V(x,\lambda) = \sum_n V_n(x)h_n(\lambda)$, where $h_n(\lambda)$ is the one-particle eigenvalue of the $n$'th charge~\cite{10.21468/SciPostPhys.2.2.014}. Also $\ell = \tfrac{L\ind{imp}}{L} = L^{1-\gamma}$ is the size of the impurity as seen from the macroscopic system. We now want to study the above equation for small $\ell \to 0$. For that let us change the position variable to $x \to x \ell$. This yields:
\begin{align}
	\ell\partial_t ({P'}\upd{dr}n) + \partial_x((\partial_\lambda E)\upd{dr}n) - \partial_\lambda((\partial_x E)\upd{dr} n) = 0,\label{equ:mesoscopic_derivation_GHD_rescaled}
\end{align}
and by taking the limit $\ell \to 0$ we find the stationary GHD equation:
\begin{align}
	\partial_x((\partial_\lambda E)\upd{dr}n) -\partial_\lambda((\partial_x E)\upd{dr} n) = 0.\label{equ:mesoscopic_derivation_GHD_stationary}
\end{align}

Note that this is a differential equation which requires boundary conditions at $x\to \pm \infty$. These are given by the asymptotic data (containing both the incoming and outgoing data) $n\ind{L}(\lambda) = n(-\infty,\lambda)$ and $n\ind{R}(\lambda) = n(\infty,\lambda)$ (Note that because we rescaled space to mesoscopic scale around the impurity, in the Euler scaling limit $x = \pm \infty$ corresponds to $x\ind{macro} = 0^{\pm}$ in macroscopic coordinates). This data has to be given externally. Again we have the problem that in order to determine which particles are incoming we need to know the complete state $n(\pm \infty,\lambda)$. This means we cannot specify the boundary conditions directly, but rather have to assume the form of $n(\pm \infty, \lambda)$ at infinity, then solve the stationary GHD equation using the incoming data and then check whether the outgoing data coincides with the incoming data.

We conclude that the impurity problem for a mesoscopic impurity reduces to the scattering problem of GHD at the impurity potential. Let us summarize some basic properties of solutions to the stationary GHD equation (\ref{equ:mesoscopic_derivation_GHD_stationary}).

\subsection{Particle number and energy conservation}

We can establish particle number conservation by integrating (\ref{equ:mesoscopic_derivation_GHD_stationary}) over $\lambda$:
\begin{align}
	\partial_x \int\dd{\lambda} (\partial_\lambda E)\upd{dr}(x,\lambda) n(x,\lambda) = 2\pi \partial_x \int\dd{\lambda} v\upd{eff}(x,\lambda) \rho(x,\lambda) = 0,\label{equ:mesoscopic_particle_number_conservation}
\end{align} 
which means that the total current $\int\dd{\lambda} v\upd{eff}(x,\lambda) \rho(x,\lambda)$ is independent of $x$ and thus the outgoing current is equal to the incoming current.

Similarly we can establish energy conservation by showing that the energy current vanishes:
\begin{align}
    &\dv{x} \int\dd{\lambda} E(x,\lambda) (\partial_\lambda E)\upd{dr}(x,\lambda)n(x,\lambda)\\
    &= \int\dd{\lambda} \partial_x E(x,\lambda) (\partial_\lambda E)\upd{dr}(x,\lambda)n(x,\lambda) + \int\dd{\lambda} E(x,\lambda) \partial_x \qty[(\partial_\lambda E)\upd{dr}(x,\lambda)n(x,\lambda)]\\
    &= \int\dd{\lambda} \partial_x E(x,\lambda) (\partial_\lambda E)\upd{dr}(x,\lambda)n(x,\lambda) - \int\dd{\lambda} \partial_\lambda E(x,\lambda) (\partial_x E)\upd{dr}(x,\lambda)n(x,\lambda) = 0,
\end{align}
where the last line vanishes due to the well-known symmetry property of the dressing $\int\dd{\lambda} fng\upd{dr} = \int\dd{\lambda} f\upd{dr}ng$ (see for instance ~\cite{BenGHD} equation (117)).

\subsection{Rescaling invariance}
\label{sec:rescaling}
There is also another immediate property following from the way we derived the stationary GHD equation: Consider a global rescaling of space $V(x,\lambda) \to V(\alpha x,\lambda)$. The solution to the stationary GHD equation in the rescaled potential is simply given by $n(\alpha x, \lambda)$. This can either be checked explicitly or can simply be observed by noting that if $\ell$ in derivation (\ref{equ:mesoscopic_derivation_GHD_rescaled}) was a mesoscopic scale then $\ell/\alpha$ is a mesoscopic scale as well and thus the resulting GHD equation and its solution must be the same, only rescaled.

This shows that the stationary GHD equation is invariant under global rescaling of space. Interestingly the same is true for local rescaling as well:

\begin{observation}
    Consider a smooth (or at least differentiable) map $y: \mathbb{R} \to \mathbb{R}$ s.t. $y(x\to \pm\infty) = \pm\infty$. Then the stationary GHD equation with potential $V(y(x),\lambda)$ is solved by $n(y(x),\lambda)$.
\end{observation}

This can be seen rather easily: We have
\begin{align}
    \partial_{\lambda} E(y(x),\lambda) &= \partial_{\lambda} E_0(\lambda) + \partial_\lambda V(y(x),\lambda)\\
    \partial_x E(y(x),\lambda) &=  \partial_xV(y(x),\lambda )y'(x).
\end{align}

Since the dressing operation does not depend on $x$ we simply find their dressed expressions evaluated at $y(x)$:
\begin{align}
    (\partial_{\lambda} E)\upd{dr} &= (\partial_{\lambda} E_0(\lambda))\upd{dr}(y(x),\lambda) + (\partial_\lambda V)\upd{dr}(y(x),\lambda)\\
    (\partial_x E)\upd{dr} &= (\partial_xV)\upd{dr}(y(x),\lambda)y'(x).
\end{align}

We also find that ${P'}\upd{dr} = {P'}\upd{dr}(y(x),\lambda)$. Now inserting everything into the stationary GHD equation we immediately find:
\begin{align}
    \qty[\partial_x((\partial_\lambda E)\upd{dr}n) -\partial_\lambda((\partial_x E)\upd{dr} n)]_{x\to y(x)}y'(x) = 0.
\end{align}

Therefore $n(y(x),\lambda)$ is a solution to the locally rescaled GHD equation. Note that $y(x)$ is not required to be monotonically increasing. It is allowed to go back in space. 

This rescaling property can be used to simplify problems. For instance, consider an impurity potential which depends only on space $V(x)$ and is non-negative and has a single maximum of $\bar{V} = \max_{x} V(x)$ which is obtained at $x_0$. Then we can start from the potential:
\begin{align}
    V\ind{tri}(x) = \bar{V}\begin{cases}1-\abs{x} & \abs{x} < 1\\
    0 & \mathrm{else}.\end{cases}
\end{align}

Now define $y(x)$ as:
\begin{align}
    y(x) &= \frac{\bar{V}-V(x)}{\bar{V}}\sgn{x-x_0}.
\end{align}

\begin{figure}[!h]
    \centering
    \includegraphics[width=\linewidth]{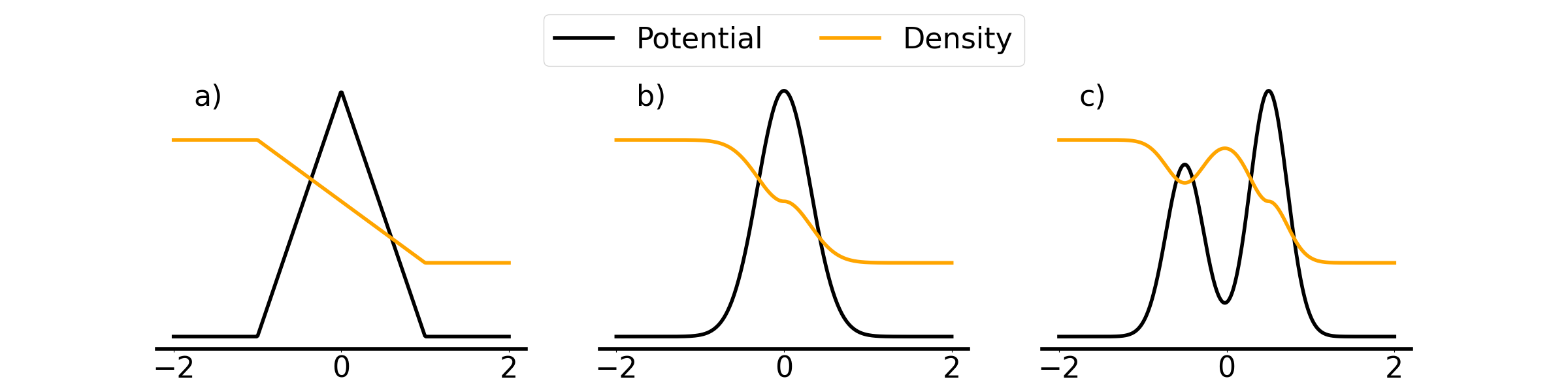}
    \caption{Schematic sketch of transformation of solution under local rescaling: a) Triangle potential and its stationary solution (sketch), b) Gaussian potential and solution obtained via rescaling of a), c) Potential with local minimum and solution obtained via rescaling of a). Note that the density increases again in the minimum. These particles there are trapped in the potential and only passively contribute to the scattering. A stationary state without trapped particles also exists, but may look very different.}
    \label{fig:mesoscopic_rescaling}
\end{figure}

Then $V\ind{tri}(y(x)) = V(x)$ and thus we can obtain the solutions to all these potentials simply by solving one potential $V\ind{tri}$. If $V(x)$ only has no other local maxima than the global one this also produces the expected result. However, if $V(x)$ has some local minima, $y(x)$ is not monotonically increasing: We still find a solution to the stationary GHD equation, but the minima are filled with some quasi-particles, which are trapped in the minima (see Figure~\ref{fig:mesoscopic_rescaling}). We usually assume the impurity to be empty before scattering, but in principle one could also consider a filled impurity.

\subsection{Impact of the sign of $T(\lambda,\mu)$}
\label{sec:mesoscopic_sign}

In general, $T(\lambda,\mu)$ describes the effective displacement on the trajectories of two quasi-particles after they scattered. If it is negative it describes a sudden forward jump of the quasi-particles during scattering (like in the hard rods case $T(\lambda,\mu)=-\tfrac{d}{2\pi}$). Positive $T(\lambda,\mu)$ (like in the Lieb-Liniger model) physically corresponds to a time-delay during scattering, but can be thought off as a backward displacement of the particles (like in the flea gas algorithm~\cite{PhysRevLett.120.045301}).

Imagine a particle moving `up-hill' at a potential $V(x)$ and consider the case where it scatters with a reflected particle going `down-hill'. If $T(\lambda,\mu)$ is negative then the two particles will be displaced forward, meaning that the incoming particle gains potential energy $\Delta E = V'(x) 2\pi |T(\lambda,\mu)|$, while the reflected particle looses this energy. Since this effectively lowers the height of the potential barrier for the incoming particle, we conclude that negative $T(\lambda,\mu)$ tends to help incoming particles to pass a potential barrier.

For $T(\lambda,\mu)>0$ the situation is reversed. Now both particles are displaced backwards, meaning the incoming particle looses potential energy, while the reflected one gains it. Therefore the potential barrier is effectively higher.

Note that this also has an impact on the uniqueness of solutions. For $T(\lambda,\mu) < 0$ reflected particles will help other particles to pass the barrier, which leads to less reflected particles. Therefore, there is a competition between the number of reflected particles and their effect on the incoming particles, which should intuitively lead to a stable equilibrium. For $T(\lambda,\mu) > 0$, however, both effects act in the same direction: If there are many reflected particles even more will get reflected. Alternatively, if only few particles are reflected more particles will be transmitted. This can lead to two solutions separated by an unstable equilibrium in between and thus the solution might not be unique.

We will see this phenomenon in the discussion of the hard rods (see Section \ref{sec:hard_rods}).

\section{Stationary GHD equation for $T(\lambda-\mu)$ and effective Hamiltonian}
\label{sec:stationary}
We already established some general properties of the solutions in the last section. Now we will specifically study the case where the scattering phase shift depends only on the difference of momenta $T(\lambda,\mu) = T(\lambda-\mu)$, where great simplifications happen since the normal modes are known.

\subsection{Stationary GHD equation in normal modes}
The stationary GHD equation (\ref{equ:mesoscopic_derivation_GHD_stationary}) is hard to solve because it is written in a continuity equation form $\partial_x (An) = \partial_\lambda (Bn)$. It would be more convenient to write the equation in transport form:
\begin{align}
    \tilde{A}\partial_x m = \tilde{B}\partial_\lambda m,
\end{align}
where $m(x,\lambda)$ are called the normal modes. In general, it is not clear how to derive the normal modes, or whether they exist. However, in the case of a scattering phase shift which only depends on the differences of the momenta $T(\lambda,\mu) = T(\lambda-\mu)$ the normal modes exists and are simply given by $n(x,\lambda)$~\cite{10.21468/SciPostPhys.2.2.014}, similar to the GHD equation without external potential. This gives the stationary GHD equation in transport form:

\begin{align}
    (\partial_\lambda E)\upd{dr} \partial_x n - (\partial_x E)\upd{dr} \partial_\lambda n = 0.\label{equ:stationary_normal_GHD}
\end{align}

For other models where $T(\lambda,\mu) \neq T(\lambda-\mu)$ this does not work. However, one could try to reparametrize $\lambda \to f(\lambda)$ or change the definition of $P(\lambda)$ (in the GHD context the momentum function $P(\lambda)$ is not physically accessible, meaning that it is some `gauge' degree of freedom~\cite{Bonnemain_2022}). Both actions change the scattering phase shift and it might be possible to obtain a $T(\lambda,\mu)$ that depends only on the difference. Alternatively, if one is still able to find some other normal modes, the derivations done in this section will still be applicable with minor adaptions.

Let us briefly recap why the occupation function $n$ are the normal modes of the GHD equation in the case $T(\lambda,\mu) = T(\lambda-\mu)$. The stationary GHD equation in conservation form is given by:
\begin{align}
    0&= \partial_x((\partial_\lambda E)\upd{dr} n) - \partial_\lambda((\partial_x E)\upd{dr} n)\\
    &= \qty[\partial_x(\partial_\lambda E)\upd{dr} - \partial_\lambda(\partial_x E)\upd{dr}] n + (\partial_\lambda E)\upd{dr} \partial_xn - (\partial_x E)\upd{dr} \partial_\lambda n.\label{equ:stationary_normal_GHD_derivation_step_1}
\end{align}

Let us write out the derivatives of the dressing explicitly:
\begin{align}
    \partial_x (\partial_\lambda E)\upd{dr} &= \partial_x\partial_\lambda E + \vu{T} \qty(\partial_xn (\partial_\lambda E)\upd{dr} + n \partial_x(\partial_\lambda E)\upd{dr})\\
    \partial_\lambda (\partial_x E)\upd{dr} &= \partial_\lambda\partial_x E + \vu{T} \qty(\partial_\lambda n (\partial_x E)\upd{dr} + n \partial_\lambda (\partial_x E)\upd{dr}).
\end{align}

Here we used $T(\lambda,\mu) = T(\lambda-\mu)$, which implies $\qty[\partial_\lambda, \vu{T}] = 0$ to swap $\partial_\lambda$ and $\vu{T}$. Using the definition of the dressing we can rewrite this as:
\begin{align}
    \partial_x (\partial_\lambda E)\upd{dr} &= \qty[\partial_x\partial_\lambda E + \vu{T} \partial_xn (\partial_\lambda E)\upd{dr}]\upd{dr}\\
    \partial_\lambda (\partial_x E)\upd{dr} &= \qty[\partial_\lambda\partial_x E + \vu{T} \partial_\lambda n (\partial_x E)\upd{dr}]\upd{dr}.
\end{align}
In particular, we find that:
\begin{align}
    \partial_x (\partial_\lambda E)\upd{dr}-\partial_\lambda (\partial_x E)\upd{dr} &= \qty[\vu{T} \qty(\partial_xn (\partial_\lambda E)\upd{dr}-\partial_\lambda n (\partial_x E)\upd{dr})]\upd{dr}\label{equ:stationary_nomal_curl}.
\end{align}

From this we can easily see that if (\ref{equ:stationary_normal_GHD}) holds (\ref{equ:stationary_nomal_curl}) and thus also (\ref{equ:stationary_normal_GHD_derivation_step_1}) will vanish. 

The advantage of (\ref{equ:stationary_normal_GHD}) over (\ref{equ:mesoscopic_derivation_GHD_stationary}) is that solutions to (\ref{equ:stationary_normal_GHD}) can be described in terms of characteristics. Consider a particle moving along a solution of the characteristic ODE:
\begin{align}
    \dv{t}x(t)&= (\partial_\lambda E)\upd{dr}(x(t),\lambda(t)) & \dv{t}\lambda(t)&= (\partial_x E)\upd{dr}(x(t),\lambda(t)).\label{equ:mesoscopic_stationary_ode}
\end{align}

Then it is easy to see that $n(x(t),\lambda(t))$ is constant in time, i.e.\ $n$ is constant along characteristics. This property is particularly useful for numerical simulations (see \ref{sec:simulation_direct}).

\subsection{Effective Hamiltonian}
Writing the stationary GHD equation in transport form is already very useful. However, it becomes even more interesting if one notices an additional fact. As a byproduct of the derivation of the stationary GHD equation in normal modes we also showed that $\partial_x (\partial_\lambda E)\upd{dr} = \partial_\lambda (\partial_x E)\upd{dr}$, see equation (\ref{equ:stationary_nomal_curl}). This implies that there exists a function $H(x,\lambda)$ with the property:
\begin{align}
    \grad H(x,\lambda) &= (\grad E)\upd{dr},
\end{align}
where $\grad = \begin{pmatrix}\partial_x & \partial_\lambda\end{pmatrix}$. We call this function $H(x,\lambda)$ the effective Hamiltonian for the following reason: The stationary GHD equation can be written as
\begin{align}
    \partial_\lambda H \partial_x n = \partial_x H \partial_\lambda n,\label{equ:stationary_effective_hamiltonian_GHD}
\end{align}
which is precisely the equation for a distribution of non-interacting particles evolving according to the Hamiltonian $H$. If there was another time derivative this equation would be the Liouville equation for that Hamiltonian. Since the time derivative is not there, this problem is now the scattering problem for non-interacting particles with Hamiltonian $H(x,\lambda)$.

This idea is typical in integrable models and in GHD. One tries to rewrite the problem in terms of another effective problem for non-interacting particles where all quantities get dressed by the other particles (consider for instance the effective velocity $v\upd{eff}(x)$ which is not the bare velocity of the particles, but gets altered due to the presence of other particles). In that respect we can view the Hamiltonian $H(x,\lambda)$ as the effective (or dressed) Hamiltonian under which the particles evolve. The effective Hamiltonian $H(x,\lambda)$ will be given by the bare Hamiltonian $E(x,\lambda)$ plus some other terms which originate from the interactions between particles.

We want to note that $H(x,\lambda)$ is technically equal to $E\upd{Dr}$ defined in~\cite[supplementary material, equation (5)]{PhysRevLett.124.140603}. However, there the authors only look at the $\lambda$ dependence of $E\upd{Dr}(\lambda)$, which works in general for any $T(\lambda,\mu)$. Since we are also interested in the $x$ dependence, the situation is more complicated and only gives (\ref{equ:stationary_effective_hamiltonian_GHD}) in the case $T(\lambda-\mu)$.

Let us now derive a formula for the effective Hamiltonian. We will start with the definition of the dressing operation:
\begin{align}
    \grad H &= \grad E + \vu{T} (n\grad H).
\end{align}

The curl of $n\grad H$ is given by:
\begin{align}
    \partial_x (n\partial_\lambda H) - \partial_\lambda(n\partial_x H) = \partial_x n \partial_\lambda H - \partial_\lambda n\partial_x H = 0,
\end{align}
which vanishes due to the stationary GHD equation. Therefore, we can find a function $N(x,\lambda)$ with the property:
\begin{align}
    \grad N = n \grad H,
\end{align}
and thus:
\begin{align}
    \grad H &= \grad E + \vu{T} \grad N.
\end{align}

Let us take this equation in $x$ and integrate it from $-\infty$ to $x$:
\begin{align}
    H(x,\lambda) &= H(-\infty,\lambda) + V(x,\lambda) + \vu{T} \qty(N(x,\lambda)-N(-\infty,\lambda)).\label{equ:stationary_effective_Hamiltonian}
\end{align}

By looking at $x \to -\infty$ we can fix $H(-\infty,\lambda)$:
\begin{align}
    H(-\infty,\lambda) &= \int_0^\lambda\dd{\lambda} (\partial_\lambda E)\upd{dr}(-\infty,\lambda) + C,
\end{align}
where we are free to choose the constant $C$.

We can interpret the effective Hamiltonian (\ref{equ:stationary_effective_Hamiltonian}) as follows: The first part $H(-\infty,\lambda)$ describes the evolution according to the GHD equation without impurity. The second term $V(x,\lambda)$ describes the bare contribution to the impurity. The third term:
\begin{align}
    U(x,\lambda) &= \vu{T} \qty(N(x,\lambda)-N(-\infty,\lambda)),\label{equ:stationary_U_def}
\end{align}
describes the contribution to the impurity coming from the interactions between particles. Note that while $V(x,\lambda)$ will vanish for large $x$, $U(x,\lambda)$ will vanish by definition for $x\to-\infty$, but in general does not need to vanish for $x \to \infty$. In fact, this is important as $U(x\to \infty,\lambda) = 0$ implies that the state on the right side of the impurity is unaffected by the impurity, i.e.\ it signals full transmission. 

As we will establish in the next section, $N(x,\lambda)$ will not vanish for large $\lambda \to \pm \infty$ but rather approaches a constant. Therefore, if $T(\lambda)$ is an integrable function, i.e.\ $\int\dd{\lambda} |T(\lambda)| < \infty$ this allows us to remove the second term in $U$ by redefining $H(-\infty,\lambda)$ and $U(x,\lambda) \to \tilde{U}(x,\lambda) = \vu{T} N(x,\lambda)$. This gives the following simplified expression for the Hamiltonian:
\begin{align}
    H(x,\lambda) &= E(x,\lambda) + \vu{T} N(x,\lambda),\label{equ:stationary_effective_Hamiltonian_integrable}
\end{align}
which agrees with (\ref{equ:stationary_effective_Hamiltonian}), up to the arbitrary constant.

However, if $T(\lambda)$ is not integrable (for instance in the hard rods case $T(\lambda) = -d$) then this definition of $U(x,\lambda)$ would be ill-defined ($\tilde{U}(x,\lambda) = \infty$) and therefore we will stick to (\ref{equ:stationary_effective_Hamiltonian}) for the following general discussion. 

Let us give two remarks about the general form of $U(x,\lambda)$: First, note that if $T(\lambda)$ is integrable and $N$ is bounded, then $U$ is actually bounded $|U(x,\lambda)| \leq 2\sup_{\lambda} N(x,\lambda) \int\dd{\mu} |T(\mu)|$. 

Second, note that $U(x,\lambda)$ is not a general function, but has to be in the image of $\vu{T}$. An important example where this is useful is the hard rods case where $U(x) = \vu{T} N(x,\lambda) = -d\int\dd{\mu}N(x,\mu)$ is a function of $x$ only. The fact that $U(x)$ depends only on $x$ simplifies the problem so much that one can actually solve the impurity problem (see Section \ref{sec:hard_rods}). 

We will finish this section with a simple, yet physically very important observation:

\begin{observation}
    At a mesoscopic impurity all particles with rapidity $\lambda$ are either transmitted or reflected with probability one.
\end{observation}

This follows directly from the existence of the effective Hamiltonian $H(x,\lambda)$. The particles have to follow deterministic trajectories. Therefore all incoming particles with a specific rapidity will follow the same trajectory and eventually leave the impurity at the same point.  

Note that this is substantially different from scattering at microscopic impurities, where particles are reflected or transmitted with a certain probability (for instance recall scattering at an rectangular potential barrier from your undergraduate quantum mechanics course). We conclude that this is a limitation of the mesoscopic impurity model and that non-deterministic scattering is an effect which appears only if the impurity is small enough. On the other hand, the amount of non-deterministic scattering can be used to quantify how well an impurity can be described by a mesoscopic impurity.

\subsection{The scattering problem in a general Hamiltonian}
\label{sec:stationary_scattering_problem}
\begin{figure}[!h]
    \centering
    \includegraphics[width=\textwidth]{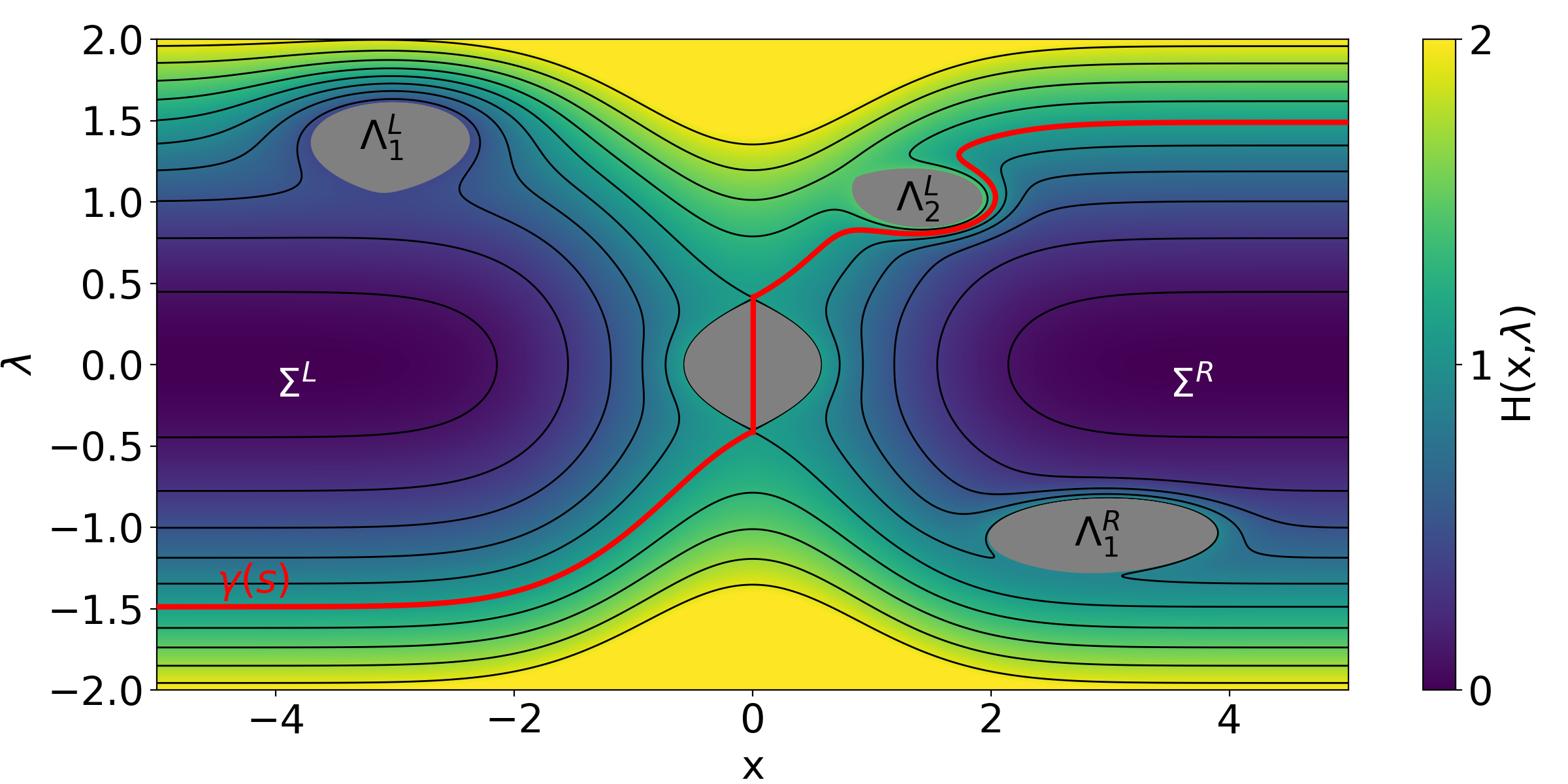}
    \caption{Contour plot of an effective Hamiltonian $H(x,\lambda)$ (for illustrative purpose only -- Hamiltonian does not correspond to a  specific system). The corresponding 3D plot is shown in Figure \ref{fig:stationary_H_3D}. The contours coincide with the trajectory of quasi-particles. We denote by $\Sigma\upd{L}$ ($\Sigma\upd{R}$) all contours corresponding to incoming particles on the left (right). Both regions are separated by the curve $\gamma(s)$ (red), which is the boundary between reflected and transmitted particles. The `islands' (gray) are regions contours whose energy is too high or too low to be reached by the incoming particles. The density of particles there is zero.}
    \label{fig:stationary_H_contour}
\end{figure}
\begin{figure}[!h]
    \centering
    \includegraphics[width=\textwidth]{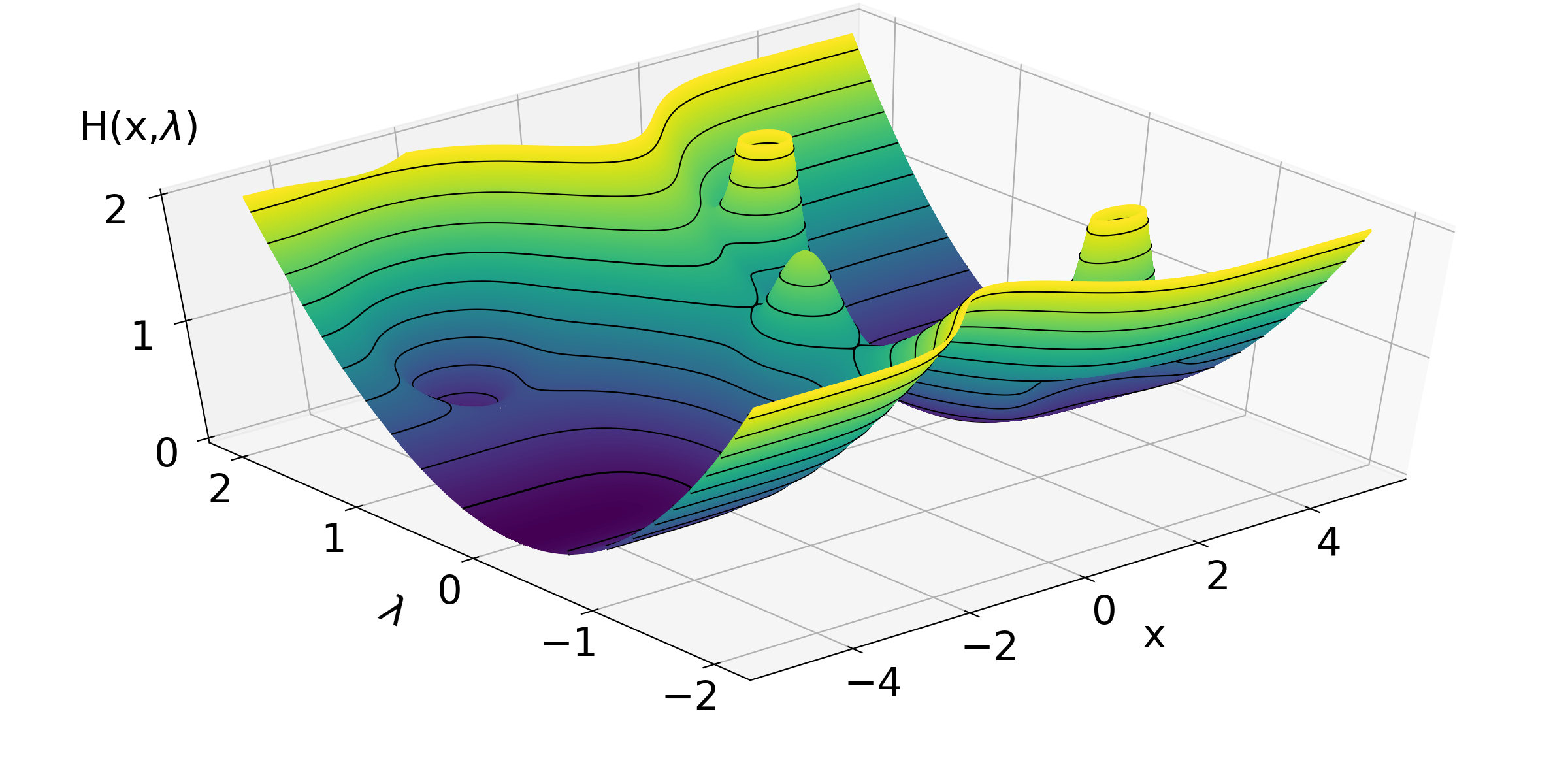}
    \caption{3D Plot of an effective Hamiltonian (corresponds to Figure \ref{fig:stationary_H_contour})}
    \label{fig:stationary_H_3D}
\end{figure}

In the previous section we introduced the effective Hamiltonian:
\begin{align}
    H(x,\lambda) = H(-\infty,\lambda) + V(x,\lambda) + U(x,\lambda),\label{equ:stationary_H_U_def}
\end{align}
with:
\begin{align}
    U(x,\lambda) &= \vu{T} \qty(N(x,\lambda)-N(-\infty,\lambda)).
\end{align}

So far this result not useful since we do not know $N(x,\lambda)$. For that we need to solve the scattering problem in the effective Hamiltonian $H(x,\lambda)$: We know that both $n(x,\lambda)$ and $N(x,\lambda)$ satisfy the following equation
\begin{align}
    \partial_\lambda H \partial_x N = \partial_x H \partial_\lambda N.
\end{align}

As we described earlier this equation implies $N(x,\lambda)$ is constant along characteristics:
\begin{align}
    \dv{t} x(t) &= \partial_\lambda H & \dv{t} \lambda(t) &= -\partial_x H.
\end{align}

However, we also know that a basic property of Hamiltonian systems is that the Hamiltonian is conserved along trajectories. This means a trajectory follows a level set of the Hamiltonian $H(x(t),\lambda(t)) = \mathrm{const}$. Now we know that $N$ and $H$ are both constant along a trajectory which implies we can locally write $N(x,\lambda) = N(H(x,\lambda))$. Locally here means that this is true for a domain $\Omega \subset \mathbb{R}^2$ which does not contain any critical points $\grad H = 0$ and does not contain more than one connected part of each level set. The reason why this is important is that globally the level sets of $H(x,\lambda)$ can have multiple disconnected components. If one starts on one of these components and follows the level set one never reaches the other disconnected components and thus there is no reason why $N(x,\lambda)$ should have the same value on these other components.

Now let us assume we know all the incoming data $n\ind{L}(\lambda) = n(-\infty,\lambda)$ and $n\ind{R}(\lambda) = n(\infty,\lambda)$ on the left and the right side. As we discussed beforehand this typically requires the full knowledge about the state including the outgoing data as well. 

For simplicity we will also restrict to the case where $(\partial_\lambda E)\upd{dr}(\pm \infty,\lambda)$ is a monotonically increasing function. This has the advantage that we can be sure that there is only one $\lambda$ which corresponds to zero velocity and on the left side all particles with higher momenta are incoming and all particles with lower momenta are outgoing (and similar on the right side).

If $(\partial_\lambda E)\upd{dr}(\pm \infty,\lambda)$ is not monotonically increasing then we might have several momenta corresponding to zero velocity which complicates the following reasoning. Nevertheless, by introducing more sets $\Sigma$, it will be possible to also include that case.

Starting from the incoming data on both sides we can first compute the Hamiltonians $H\upd{L/R}(\lambda)$ on both sides (which we know up to a constant) and then we can define the two functions:
\begin{align}
    \tilde{N}\upd{L}(h) &= \int_{H\upd{L}(\lambda\ind{L})}^h\dd{h} n\ind{L}(H^{L^{-1}}(h))\\
    \tilde{N}\upd{R}(h) &= \int_{H\upd{R}(\lambda\ind{R})}^h\dd{h} n\ind{R}(H^{R^{-1}}(h)),
\end{align}
where $\lambda\ind{L/R}$ are the momenta corresponding to zero velocity on both sides and $H^{L/R^{-1}}(h)$ is the inverse function of $H\upd{L/R}(\lambda)$ where we choose the incoming branch, i.e.\ $H^{L^{-1}}(h) > \lambda\ind{L}$ and $H^{R^{-1}}(h) < \lambda\ind{R}$.\\

We will now describe how to construct $N(x,\lambda)$ from the incoming data in a general Hamiltonian. This construction is also illustrated in Figure \ref{fig:stationary_H_contour} for the example Hamiltonian depicted in Figure \ref{fig:stationary_H_3D}.

Given a general Hamiltonian we can group its level sets into three type of sets: First, we define $\Sigma\upd{L/R}$ which are all the level sets that originate from incoming data on either the right or the left side. We think of this as the regions which are allowed for particles to enter. For instance, $\Sigma\upd{L}$ consist of level sets which start from the $x = -\infty$ on the left side with positive velocity and then either extend all the way to either the right side (i.e.\ the particle is transmitted) or it bends back and extends to the left side again (i.e.\ the particle is reflected). Then we have a collection of sets $\Lambda\upd{L/R}_i$ which are `islands' in either $\Sigma\upd{L/R}$ where the energy $H(x,\lambda)$ is too high (or to low) for particles to enter. $\Lambda\upd{L/R}_i$ consist of closed loops of level sets. Note that we associate to each of these $\Lambda\upd{L/R}_i$ an energy $h\upd{L/R}_i = H(\partial \Lambda\upd{L/R}_i)$ which corresponds to the energy of particles `flowing' around $\Lambda\upd{L/R}_i$. The last type of sets are non-allowed regions similar to the $\Lambda\upd{L/R}_i$ which lie between $\Sigma\upd{L}$ and $\Sigma\upd{R}$. Again they consist of closed loops of level sets. However, unlike the $\Lambda\upd{L/R}_i$ these islands will not give a contribution as we will define $N(x,\lambda) = 0$ on them. 

The two sets $\Sigma\upd{L/R}$ are separated by a curve $\gamma(s) \in \mathbb{R}^2$. This curve might not necessarily be unique, it is only required that $\Sigma\upd{L}$ lies on one side of the curve and $\Sigma\upd{R}$ on the other. We will parametrize this curve in such a way that $\gamma_x(s \to \pm\infty) = \pm \infty$. In many cases the curve $\gamma(s)$ will be a function of $x$ only in which case we write $\gamma = (x,\lambda_0(x))$. Note that $\lambda_0(-\infty) = \gamma_\lambda(-\infty)$ has the interpretation of the largest rapidity incoming from the left side which is reflected by the impurity and the smallest rapidity coming from the right side which is transmitted by the impurity (and similarly for $\lambda_0(\infty) = \gamma_\lambda(\infty)$ on the right side). We will fix the integration constants in such a ways that both $N(\gamma(s)) = 0$ and the Hamiltonian $H(\gamma(s)) = 0$ on this curve. 

Once these sets have been characterized one can explicitly define $N(x,\lambda)$:
\begin{align}
    N(x,\lambda) &= \begin{cases}N\upd{L}(H(x,\lambda)) = \tilde{N}\upd{L}(H(x,\lambda)-H(-\infty,\lambda)+H\upd{L}(\lambda))-\tilde{N}\upd{L}(H\upd{L}(\lambda_0(-\infty)))\\
    \hspace{10cm} (x,\lambda) \in \Sigma\upd{L}\\
    N\upd{R}(H(x,\lambda)) = \tilde{N}\upd{R}(H(x,\lambda)-H(\infty,\lambda)+H\upd{R}(\lambda))-\tilde{N}\upd{R}(H\upd{R}(\lambda_0(\infty)))\\
    \hspace{10cm} (x,\lambda) \in \Sigma\upd{R}\\
    N(\partial \Lambda\upd{L/R}_i) = N\upd{L/R}(h\upd{L/R}_i)\\
    \hspace{10cm} (x,\lambda) \in \Lambda\upd{L/R}_i.
    \end{cases}\label{equ:stationary_N_formula}
\end{align}

The first two lines just copy the incoming data into the regions $\Sigma\upd{L/R}$. For instance, on $\Sigma\upd{L}$ we know that $N(x,\lambda) = \tilde{N}\upd{L}(H(x,\lambda)+ C) + D$, where $C$ and $D$ are constants we are allowed to choose. We set $C = H\upd{L}(\lambda)-H(-\infty,\lambda)$ and $D=\tilde{N}\upd{L}(H\upd{L}(\lambda_0(-\infty)))$, to ensure that $N(\gamma(s)) = H(\gamma(s)) = 0$. In the non-allowed regions $\Lambda\upd{L/R}_i$ we know that $n(\Lambda\upd{L/R}_i) = 0$ and thus $N(\Lambda\upd{L/R}_i) = \mathrm{const}$. The constant is determined by its value on the boundary $\partial \Lambda\upd{L/R}_i$, which corresponds to a level set of the Hamiltonian with energy $h\upd{L/R}_i$. 

This complicated procedure allows us to compute $N(x,\lambda)$ given $H(x,\lambda)$. In turn we can now use this to compute $U(x,\lambda)$: Let us define $\Sigma\upd{L/R}(x) = \qty{\lambda:(x,\lambda)\in \Sigma\upd{L/R}}$ and $\Lambda\upd{L/R}_i(x) = \qty{\lambda:(x,\lambda)\in \Lambda\upd{L/R}_i}$.

Then we can rewrite definition (\ref{equ:stationary_U_def}) as:
\begin{align}
    U(x,\lambda) = \sum_{r\in\qty{L,R}}\Big[&\int_{\Sigma\upd{r}(x)}\dd{\mu}T(\lambda-\mu) N\upd{r}(H(x,\mu)) + \sum_i \int_{\Lambda\upd{r}_i(x)}\dd{\mu}T(\lambda-\mu) N\upd{r}(h\upd{r}_i)\nonumber\\
    &- \int_{\Sigma\upd{r}(-\infty)}\dd{\mu} T(\lambda-\mu) N\upd{r}(H(-\infty,\mu))\Big].
\end{align}

Note that if the $T(\lambda)$ is not integrable, then the individual integrals might not converge since $N(\lambda\to\infty) \to \mathrm{const}$. In this case one has to combine the integrals into one integral over the sum of all three terms (which will then be finite since all constants at infinity cancel). We can also insert expression (\ref{equ:stationary_H_U_def}) for $H(x,\lambda)$:
\begin{align}
    U(x,\lambda) = G[U](x,\lambda) = \sum_{r\in\qty{L,R}}\Big[&\int_{\Sigma\upd{r}(x)}\dd{\mu}T(\lambda-\mu) N\upd{r}(H(-\infty,\mu) + V(x,\mu) + U(x,\mu))\nonumber\\
    &+ \sum_i \int_{\Lambda\upd{r}_i(x)}\dd{\mu}T(\lambda-\mu) N\upd{r}(h\upd{r}_i)\nonumber\\
    &- \int_{\Sigma\upd{r}(-\infty)}\dd{\mu} T(\lambda-\mu) N\upd{r}(H(-\infty,\mu))\Big].\label{equ:stationary_fixed_point}
\end{align}

This is a functional fixed point problem for $U(x,\lambda)$ with fixed point functional $G[U]$. Note that this fixed point problem is quite complicated, in particular since the shape of the domains $\Sigma\upd{L/R}$ and $\Lambda\upd{L/R}_i$ depend on $U(x,\lambda)$. In case we manage to find a solution to the fixed point equation we can write the solution to the stationary GHD equation (\ref{equ:stationary_normal_GHD}) as follows:
\begin{align}
    n(x,\lambda) &= \begin{cases}n\ind{L}(H^{L^{-1}}(H(x,\lambda)-H(-\infty,\lambda)+H\upd{L}(\lambda))) & (x,\lambda) \in \Sigma\upd{L}\\
    n\ind{R}(H^{R^{-1}}(H(x,\lambda)-H(-\infty,\lambda)+H\upd{R}(\lambda))) & (x,\lambda) \in \Sigma\upd{R}\\
    0 & (x,\lambda) \in \Lambda\upd{L/R}_i.
    \end{cases}\label{equ:stationary_n_solution}
\end{align}

\noindent Remark: The fixed point problem also has contributions from the regions $\Lambda\upd{L/R}_i$, even though there $n(x,\lambda) = 0$. This is because the defining relation $\grad N = n\grad H = 0$ does not not imply that $N(x,\lambda) = 0$, but only that $N(x,\lambda) = \mathrm{const}$ in $\Lambda\upd{L/R}_i$. This is mathematically similar to the `Aharonov–Bohm effect' in quantum mechanics~\cite{doi:10.1142/S0129055X15300010}.

\subsection{Discussion of the fixed point equation}

The fixed point equation (\ref{equ:stationary_fixed_point}) is quite complicated and can in general only be solved numerically. However, even without knowing the solution explicitly one can still establish some basic properties.

\subsubsection{Existence of solution}

It is physically clear that there should be some solution to the scattering problem given the incoming data. However, as we discussed earlier it is not clear how to specify the incoming data in interacting models and therefore it is also not clear whether a solution will always exits. In the following we will give a mathematical argument why we expect that the fixed point equation should always have solution for given incoming data $N\upd{L/R}$. Let us rewrite this as a functional equation:  
\begin{align}
    F[U](x,\lambda) = U(x,\lambda)- G[U](x,\lambda) = 0.
\end{align}

First, observe that this functional changes continuously under smooth deformations of $U(x,\lambda)$. This is obvious as long as changing $U$ does not create or destroy domains $\Lambda\upd{L/R}_i$ or change their energies $h\upd{L/R}_i$. Then it only shifts around the boundaries in a continuous fashion which gives a continuous change of the functional $F[U]$. Also in the other cases $F$ is continuous. For instance, when the change in $U$ creates a new domain $\Lambda\upd{L/R}_i$ we tear a `hole' into $\Sigma\upd{L/R}$ and then fill the interior with a constant value of $N$ s.t. the function $N$ is still continuous. This operation also only changes $N(x,\lambda)$ continuously and thus $F[U]$ is continuous.  

Now consider a bump function $\Delta U(x,\lambda)$ with compact support and let us study how $\int\dd{x}\dd{\lambda}F[U+\alpha \Delta U](x,\lambda)\Delta U(x,\lambda)$ behaves as $\alpha\to \pm \infty$. In both cases it is easy to see that 
\begin{align}
    \int\dd{x}\dd{\lambda}F[U+\alpha \Delta U](x,\lambda)\Delta U(x,\lambda) \to \pm \infty.\label{equ:stationary_existence_step}
\end{align}

This is particularly obvious if $T(\lambda)$ is integrable as the interaction induced term is then bounded. But also when $T(\lambda)$ is not integrable the interaction induced term will still approach a finite value as $\alpha \to \pm\infty$. No matter the sign of $\alpha$ as $|\alpha| \to \infty$ we have that $|U(x,\lambda) + \alpha \Delta U(x,\lambda)|$ will be very large around the support of $\Delta U$ and thus we have an island $\Lambda\upd{L/R}_i$ formed around the support of $\Delta U$. The value of $N(x,\lambda)$ on this island is constant and will approach $N(\partial \mathrm{supp}(\Delta U))$ which is independent of $\alpha$ for large $|\alpha|$. Thus, as $\alpha \to \pm\infty$, $N(x,\lambda)$ approaches a finite function and therefore the interaction induced term is bounded in $\alpha$, which implies (\ref{equ:stationary_existence_step}).

We have now shown that for any $\Delta U$ we have $\int\dd{x}\dd{\lambda}F[U+\alpha \Delta U](x,\lambda)\Delta U(x,\lambda) \to \pm \infty$. Let us take the formal limit where $\Delta U$ approaches a delta function which gives:
\begin{align}
    F[U+\lambda \delta(\cdot-x)\delta(\cdot-\lambda)](x,\lambda) \to \pm \infty.\label{equ:stationary_existance_delta}
\end{align}

Of course the Hamiltonian $H(x,\lambda)+\alpha \delta(\cdot-x)\delta(\cdot-\lambda)$ does not make proper sense which is why we choose to regularize them by considering bump functions. We now view $U(x,\lambda)$ as a vector $\va{U}$ and $\delta(\cdot-x)\delta(\cdot-\lambda)$ as a basis of this functional vector space. A finite-dimensional version of statement (\ref{equ:stationary_existance_delta}) is:
\begin{align}
    F_k(\va{U} + \alpha \va{e}_k) \to \pm \infty,\label{equ:stationary_existance_finite_dim}
\end{align}
for all $k$ as $\alpha \to \infty$. If (\ref{equ:stationary_existance_finite_dim}) holds in one dimension and $F$ is continuous then the intermediate value theorem shows that there must be a zero of $F(U)$ on the real line. In the finite-dimensional case there is a generalization of the intermediate value theorem, called Poincaré–Miranda theorem, which (up to some mathematical technicalities) establishes that (\ref{equ:stationary_existance_finite_dim}) implies the existence of a zero of $F[U]$. This indicates that (\ref{equ:stationary_existance_delta}) should imply the existence of a solution $F[U] = 0$ as well (however, this is not rigorous as results in finite dimension possibly do not carry over to the infinite-dimensional setting). 

\subsubsection{Discussion of uniqueness}
\label{sec:stationary_uniqueness}
While the existence of solutions is physically expected, there is no reason why solutions should be unique. Non-uniqueness of solutions means that the solution will depend on the scattering history. Therefore studying the uniqueness of solutions gives physical insights. For that it is constructive to compute the Jacobian of $F[U]$, i.e.\ how $F$ is perturbed by small perturbations of $U(x,\lambda) \to U(x,\lambda) + \delta U(x,\lambda)$:

Let us again consider a $\delta U$ which is a bump function. The important observation here is that as long as $\delta U$ does not affect any critical points of $H(x,\lambda)$ (i.e.\ it does not move them around or create/destroy them) the boundaries of the sets $\Sigma\upd{L/R}$ and $\Lambda\upd{L/R}_i$ might change in a continuous fashion, but the topology of the sets does not change. In particular the energy levels $h\upd{L/R}_i$ associated to $\Lambda\upd{L/R}_i$ remain unchanged. For such a perturbation we therefore find:

\begin{align}
    \qty(JF[U])\delta U &= \dv{\alpha}F[U+\alpha \delta U]\eval_{\alpha=0}\\
    &= \delta U(x,\lambda)\nonumber \\
    &\hspace{0.5cm}-\sum_{r\in\qty{L,R}}\Big[\int_{\Sigma\upd{r}(x)}\dd{\mu}T(\lambda-\mu) n\upd{r}(H(-\infty,\mu) + V(x,\mu) + U(x,\mu))\delta U(x,\mu)\Big]\\
    &= (1-\vu{T}n) \delta U,
\end{align}
where we used (\ref{equ:stationary_n_solution}) to identify the solution $n(x,\lambda)$. Again we let $\delta U(x,\lambda) \to \delta(x-x_0)\delta U(\lambda)$ approach a delta function in space which gives:
\begin{align}
    \qty(JF[U])\qty(\delta(x-x_0)\delta U) = \delta(x-x_0)(1-\vu{T}n(x_0))\delta U,
\end{align}
where $n(x_0,\lambda)$ is the density at $x_0$. From this we can see that perturbations at $x_0$ only affect the result at $x_0$ but not at any other $y \neq x_0$. This means that the fixed point equation decouples for different $x$ (as long as there are no critical points at $x$). 

Uniqueness of the fixed point equation is now linked to whether the Jacobian $1-\vu{T}n(x_0)$ is invertible. We can identify this operator as the inverse dressing operator (\ref{equ:general_dressing}), which we know should be invertible if $n$ represents a physical state. Therefore, as long as we stay inside the set of physical states if we find a solution at $x_0$ it will be a unique one. There can be multiple solutions but they either need to be separated by non-physical states or by changes of the critical points. In particular quantum mechanical models (like the Lieb-Liniger model) where $T(\lambda)$ is integrable $\int\dd{\lambda} T(\lambda) = 1$ and in addition the occupation function is bounded $n(x,\lambda) < 1$ (i.e.\ every quantum number can be occupied at most once) the Jacobian will always be invertible (in that case the operator norm $\norm{\vu{T}n}_\infty < 1$ and it is a standard result that $1-\vu{T}n$ is invertible).

The above arguments only partially answer the question of uniqueness since there we excluded the case when $\delta U$ affects critical points of the Hamiltonian. It only tells us that the Jacobian is invertible on that subspace. We also need to discuss what happens in case $\delta U$ affects a critical point. This is hard to analyze in general since changing $U$ at a critical point will affect the solution globally. For instance, when $\delta U$ acts on a critical point on the boundary of a $\Lambda\upd{L/R}_i$ it will change the energy $h\upd{L/R}_i$ associated to $\Lambda\upd{L/R}_i$, which (since the boundary is a level set of $H(x,\lambda)$ with that energy $h\upd{L/R}_i$) will move the whole boundary $\partial\Lambda\upd{L/R}_i$. 

However, the set of critical points of the Hamiltonian will usually be finite and determine the topology of the problem. If we know the positions of the critical points, the values $\lambda_0(\pm\infty) = \gamma_\lambda(\pm\infty)$ and the energies $h\upd{L/R}_i$ associated to the islands we can construct the sets $\Sigma\upd{L/R}$ and $\Lambda\upd{L/R}_i$. In a model where we can ensure that the dressing is always well defined, this finite amount of information is enough to construct a full unique solution $U(x,\lambda)$ from it. In practice this allows us to simplify the functional fixed point problem into a finite-dimensional fixed point problem, which is much easier to analyze. We will use this strategy later in the case of the hard rods model to show that for positive rod length the solution is unique, while for negative rod length we give a specific example where the solution is not unique, see Section \ref{sec:hard_rods_hysteris}. 

Remark: The discussion of uniqueness here considers only the uniqueness of the fixed point problem. The construction of the fixed point problem requires the knowledge about which particles are incoming. As we already discussed one can only determine this if one knows the outgoing particles as well. Thus, even if the fixed point problem always has a unique solution there could be multiple configurations of incoming and outgoing particles. 

\subsubsection{Solution for weak potentials}
\label{sec:weak}
Although the fixed point equation cannot be solved explicitly for a general model it is a non-perturbative method. A different, perhaps more standard, way of approaching the impurity problem would be to treat the impurity via perturbation theory in impurity strength $\alpha$. In general, the problem with perturbation theory is that it is often not clear whether the perturbative expansion indeed converges to the actual solution, or whether it misses non-perturbative effects. In this section we establish that for small impurity strength $\alpha$ the scattering will vanish identically. This implies that a perturbative expansion in $\alpha$ would yield $0$ to all orders. We conclude that scattering at mesoscopic impurities is fully non-perturbative in nature.

\begin{restatable}{observation}{obsweak}
Consider a state $n(\lambda)$ s.t. $n(\lambda)$ is identically zero in a finite region around the zero velocity rapidity, i.e.\ the density is only non-zero for $|\partial_\lambda E_0| > C$. Consider any (bounded) potential $V(x,\lambda)$ and scale it $\alpha V(x,\lambda)$ with $\alpha \to 0$. Then there exists a $\alpha_0>0$ s.t. for all $|\alpha| < \alpha_0$ there exists a solution to (\ref{equ:stationary_fixed_point}) where the states on the left and right side coincide. This means that all particles are unaffected by the impurity and no particles are reflected.
\end{restatable}

Note that this is clear for an impurity for free particles: For instance, consider the impurity $V(x,\lambda) = \bar{V}e^{-\tfrac{x^2}{2}}$. Then all particles with rapidity $\lambda > \sqrt{2\alpha\bar{V}}$ will be transmitted, while all particles with rapidity $\lambda<\sqrt{2\alpha\bar{V}}$ will be reflected. If all particles have a rapidity bounded from below $|\lambda|>\lambda_0$, then for $\alpha < \alpha_0 = \frac{\lambda_0^2}{2\bar{V}}$ all particles are transmitted. 

In interacting models the idea is similar: The interaction changes the Hamiltonian, but if $\alpha$ is small enough this change is negligible and thus the free particle result applies as well. While this establishes that particles are not reflected, it does not explain why the transmitted particles are also unaffected. For instance, the momenta of the outgoing particles could be shifted. This can only be checked by explicitly constructing the solution, which we do in \ref{sec:weak_proof}.

\section{The hard rod case}
\label{sec:hard_rods}
We would like to show how how the ideas from the last section can be applied to a specific model: the hard rods model. In this model we have $\varphi(\lambda-\mu) = -d, P(\lambda) = \lambda$ and $E_0(\lambda) = \tfrac{\lambda^2}{2}$. The physical hard rods model is given by positive rod length $d>0$, but we will also allow $d<0$. In negative hard rods the $d$ does not describe the rods size, but should rather be interpreted as the time delay for the scattering of two particles, which leads to an effective negative position shift $-d$. 

Before we start let us note a speciality of the hard rods model. One can explicitly evaluate the dressing (\ref{equ:general_dressing})~\cite{Doyon_2017,10.21468/SciPostPhys.3.6.039}:
\begin{align}
    f\upd{dr}(\lambda) = f(\lambda)-d\int\dd{\mu} f(\mu) \rho(\mu) = f(\lambda)-\tfrac{d}{2\pi}\int\dd{\mu} \tfrac{n(\mu)f(\mu)}{1+\tfrac{d}{2\pi}\int\dd{\mu}n(\mu)},
\end{align}
and the effective velocity:
\begin{align}
    v\upd{eff}(\lambda) = \frac{\lambda-d\int\dd{\mu}\mu\rho(\mu)}{1-d\int\dd{\mu}\rho(\mu)} = \qty(1+\tfrac{d}{2\pi}\int\dd{\mu}n(\mu))\lambda - \tfrac{d}{2\pi}\int\dd{\mu} \mu n(\mu),\label{equ:hard_rods_eff_velo}
\end{align}
which in particular means that the rapidity that corresponds to zero velocity is $d\expval{\lambda} := d\int\dd{\lambda} \lambda \rho(x,\lambda)$.
Using $\lambda\upd{dr} = \lambda-d\expval{\lambda} = \partial_\lambda H(\pm\infty,\lambda)$ we can explicitly compute the effective Hamiltonian outside the impurity:
\begin{align}
    H(\pm\infty,\lambda) = \frac{(\lambda-d\expval{\lambda})^2}{2} + C_{\pm}\label{equ:hard_rods_outside_H},
\end{align}
where $C_{\pm}$ are the integration constants on both sides. 

Note that $d\expval{\lambda}$ has the same value on both sides since the total current satisfies $\int\dd{\lambda} v\upd{eff}(\lambda) \rho(\lambda) = \int\dd{\lambda} \lambda \rho(\lambda) = \expval{\lambda}$ and the total current is independent of $x$ (see equation (\ref{equ:mesoscopic_particle_number_conservation})).

Due to $T(\lambda-\mu) = -\tfrac{d}{2\pi}$ we find that $U(x,\lambda) = U(x)$ of (\ref{equ:stationary_fixed_point}) is a function of $x$ only. Compared to the general case, where $U(x,\lambda)$ can depend on both $x$ and $\lambda$ this already simplifies the problem. We will now argue that one can simplify the fixed point equation even further and finally arrive at a finite-dimensional fixed point problem:

First, let us recall that $U(x)$ satisfies the following fixed point equation:

\begin{align}
    U(x) = G_x(U(x)) = -\tfrac{d}{2\pi}\sum_{r\in\qty{L,R}}\Big[&\int_{\Sigma\upd{r}(x)}\dd{\mu}N\upd{r}(H(-\infty,\mu) + V(x,\mu) + U(x))\nonumber\\
    &+ \sum_i \int_{\Lambda\upd{r}_i(x)}\dd{\mu} N\upd{r}(h\upd{r}_i) - \int_{\Sigma\upd{r}(-\infty)}\dd{\mu} N\upd{r}(H(-\infty,\mu))\Big].\label{equ:hard_rods_fixed_point}
\end{align}

As we discussed before (see Section \ref{sec:stationary_uniqueness}), unless at $x$ there is a critical point of $H(x,\lambda)$, then a change of $U(x)$ will only affect the fixed point equation at $x$, but not at any other position. Therefore fixing one of those $x$, where is no critical point of $H(x,\lambda)$ we can rewrite the fixed point equation as:
\begin{align}
    F_x(U) = U + G_x(U) = 0,
\end{align}
which is a function of one variable only. Furthermore by taking the derivative of $F_x$ we find for $d > 0$:
\begin{align}
    \dv{F_x(U)}{U} = 1 + \tfrac{d}{2\pi}\sum_{r\in\qty{L,R}}\Big[&\int_{\Sigma\upd{r}(x)}\dd{\mu}n\upd{r}(H(-\infty,\mu) + V(x,\mu) + U)\Big] \geq 1,\label{equ:hard_rods_F_x}
\end{align}
and therefore $F_x(U)$ is monotonically increasing. Here $n\upd{L/R}(h) = \dv{N\upd{L/R}(h)}{h}$, which we identify with the solution (\ref{equ:stationary_n_solution}), i.e.\ $n(x,\lambda) = n\upd{L/R}(h(x,\lambda))$ on $\Sigma\upd{L/R}$. This is implies uniqueness of a solution $U = U(x)$. Given the location of the critical points we can therefore compute $U(x)$ at all other positions $x$. The only missing pieces of information are locations of the critical points, $d\expval{\lambda}$ and the integration constants $C_\pm$ from (\ref{equ:hard_rods_outside_H}). Note that this is a finite set of information. Therefore, it should be possible to reduce the fixed point equation to a finite-dimensional problem. This is an abstract statement, but we will make the ideas more explicit at an example.

Also note that the above discussion was for $d>0$, for negative $d$ there is no mathematical reason why (\ref{equ:hard_rods_F_x}) should be positive. Therefore, there could be potentially more than one solution.

\subsection{Explicit example: Potential does not depend on $\lambda$ and incoming particles from the left}
\label{sec:hard_rods_example}
In the following we would like to explain this procedure on the most simple example: An impurity $V(x)$ which only depends on position $x$. For simplicity let us also restrict to $V(x) \geq 0$ which has one unique maximum $\bar{V}$ at $x = 0$ and $V'(x) > 0$ for $x<0$ and $V'(x)<0$ for $x<0$. Note that due to the local rescaling invariance all of these impurities will give rise to the same scattering (see Section \ref{sec:rescaling}). To make the analysis even more simpler let us now look at the situation where there are no incoming particles from the right, i.e.\ $n\ind{R}(\lambda) = 0$. This example can easily be extended to the case where particles also come in from the right, but then the discussion of uniqueness becomes much more complicated.

Since there are only particles coming in from the left it is convenient to redefine the zero value of the Hamiltonian to be at the zero-velocity rapidity $d\expval{\lambda}$ at $x = -\infty$. This means that the Hamiltonian is given by:
\begin{align}
    H(x,\lambda) = \tfrac{1}{2}(\lambda-d\expval{\lambda})^2 + V(x) + U(x) = \tfrac{1}{2}(\lambda-d\expval{\lambda})^2 + W(x),\label{equ:hard_rods_position_impurity_H}
\end{align}
where we combined $W(x) = V(x) + U(x)$. To understand the simplicity of the Hamiltonian (\ref{equ:hard_rods_position_impurity_H}) is, the reader might find it helpful to define $v = \lambda-d\expval{\lambda}$ and write the Hamiltonian as:
\begin{align}
    H(x,v) = \tfrac{1}{2}v^2 + W(x).
\end{align}

This is just the Hamiltonian of a classical particle moving in the effective potential $W(x)$. Scattering at such a potential is very simple. Let us denote by $\bar{W} = \max_x W(x)$ the highest value of the effective potential. As we will observe later this maximum is at $x=0$ which is the same position as the maximum of $V(x)$. Then all incoming particles with kinetic energy smaller than $\bar{W}$ will get reflected and all other particles will get transmitted.

This can be expressed in the following way (note that we slightly redefines $N(x,\lambda)$ due to the redefinition of $H(x,\lambda)$):
\begin{align}
    N(x,\lambda) = \begin{cases}
        N\upd{L}(H(x,\lambda)) & \lambda > \lambda_0(x)\\
        N\upd{L}(\bar{W}) & \lambda < \lambda_0(x).
    \end{cases}
\end{align}

Here $N\upd{L}$ and the boundary $\lambda_0(x)$ are explicitly given by:
\begin{align}
    N\upd{L}(h) &= \int_{d\expval{\lambda}}^{d\expval{\lambda}+\sqrt{2h}}\dd{\lambda} (\lambda-d\expval{\lambda}) n\ind{L}(\lambda) = \int_{0}^{\sqrt{2h}}\dd{v} v n\ind{L}(d\expval{\lambda}+v)\\ 
    \lambda_0(x) &= d\expval{\lambda} +\sgn{x} \sqrt{2(\bar{W}-W(x))}.
\end{align}

Note that in the region $\lambda<\lambda_0(x)$ there are no particles, but still $N(x,\lambda)$ is non-zero since $N(x,\lambda)$ has to be a continuous function.

From the fixed point equation for $U(x)$ we can write down a equation for $W(x)$:
\begin{align}
    F_x(W(x)) = W(x) + \tfrac{d}{2\pi} \Big[&\int_{\lambda_0(x)}^\infty\dd{\lambda} N\upd{L}(\tfrac{1}{2}(\lambda-d\expval{\lambda})^2 + W(x))\nonumber\\
    &- \int_{\lambda_0(-\infty)}^\infty\dd{\lambda} N\upd{L}(\tfrac{1}{2}(\lambda-d\expval{\lambda})^2)\nonumber\\
    &+ (\lambda_0(x)-\lambda_0(-\infty))N\upd{L}(\bar{W})\Big] = V(x).\label{equ:hard_rods_position_impurity_W_equation}
\end{align}

The last term comes from the $-N(-\infty,\mu)$ term in (\ref{equ:stationary_U_def}). Note that since $N(x,\lambda)$ is a non-zero constant for $\lambda<\lambda_0(x)$ both integrals in (\ref{equ:stationary_U_def}) are formally infinite, however the divergent part in both integrals cancels. Equation (\ref{equ:hard_rods_position_impurity_W_equation}) is particularly useful since $V(x)$ only appears on the right hand side.

Let us take the derivative of $F_x(W)$:
\begin{align}
    \dv{F_x(W)}{W} &= 1 + \tfrac{d}{2\pi} \int\dd{\lambda} n\upd{L}(\tfrac{1}{2}(\lambda-d\expval{\lambda})^2 + W(x)),\label{equ:hard_rods_example_F_x}
\end{align}
where we defined $n\upd{L}(h) = \dv{N\upd{L}(h)}{h} = n\ind{L}(d\expval{\lambda}+\sqrt{2h})$. For $d>0$ this derivative is always positive, for $d<0$ it might become negative as well. However, note that for a physical state $\dv{F_x(W)}{W}$ coincides with $\tfrac{1}{1\upd{dr}(x)}$, which has to be positive. 

From equation (\ref{equ:hard_rods_example_F_x}) we can compute the derivative of $W(x)$ w.r.t. $x$:
\begin{align}
    \dv{W(x)}{x} = \frac{1}{\dv{F_x(W)}{W}} \dv{V(x)}{x} = 1\upd{dr}(x)V'(x) = \frac{V'(x)}{1+\tfrac{d}{2\pi}\int\dd{\lambda}n(x,\lambda)}.\label{equ:hard_rods_position_impurity_dW_dx}
\end{align}

This equation has a nice interpretation: The slope of the effective potential (i.e.\ the force) is the slope of the bare potential, but modified due to the interaction with other particles. In particular for $d>0$ the slope is decreased (the potential barrier is lowered, thus more particles will be transmitted), while for $d<0$ the slope is increased (the potential barrier is higher, thus less particles will be transmitted). This is in line with the intuitive picture we had discussed before in Section \ref{sec:mesoscopic_sign}. From equation (\ref{equ:hard_rods_position_impurity_dW_dx}) we can also infer that the highest point of the effective potential $W(x)$ is at the highest point $x=0$ of $V(x)$ as well.

Let us evaluate (\ref{equ:hard_rods_position_impurity_W_equation}) at $x=0$:
\begin{align}
    F_0(\bar{W}) = \bar{W} + \tfrac{d}{2\pi} \Big[&\int_{d\expval{\lambda}}^\infty\dd{\lambda} N\upd{L}(\tfrac{1}{2}(\lambda-d\expval{\lambda})^2 + \bar{W})\nonumber\\
    &- \int_{d\expval{\lambda}-\sqrt{2\bar{W}}}^\infty\dd{\lambda} N\upd{L}(\tfrac{1}{2}(\lambda-d\expval{\lambda})^2)+ \sqrt{2\bar{W}}N\upd{L}(\bar{W})\Big] = \bar{V}.\label{equ:hard_rods_example_F_0}
\end{align}

Since $F_0$ is continuous and $F_0(0) = 0$ and $F_0(\bar{W} \to \infty) \to \infty$ we know that there will exist at least one solution $\bar{W}>0$ for $\bar{V}>0$. Again let us compute the derivative:
\begin{align}
    \dv{F_0(\bar{W})}{\bar{W}} = 1 + \tfrac{d}{2\pi} \Big[&\int_{d\expval{\lambda}}^\infty\dd{\lambda} n\upd{L}(\tfrac{1}{2}(\lambda-d\expval{\lambda})^2 + \bar{W}) + \sqrt{2\bar{W}}n\upd{L}(\bar{W})\Big].
\end{align}

This is always positive for $d>0$ and thus a unique solution exists. For $d<0$ this is not clear: While $\int\dd{\lambda} n(x,\lambda)<\tfrac{2\pi}{|d|}$ is bounded for physical reasons ($1\upd{dr}$ has to be positive) there is no bound on the second term $n\upd{L}(\bar{W}) = n\ind{L}(d\expval{\lambda}+\sqrt{2\bar{W}})$. It is not hard to construct a situation where there are multiple solutions $\bar{W}$ for some value $\bar{V}$.

Recall, however, that this solution is only a solution to the fixed point equation, which is not the full solution to the scattering problem (see the discussion at the end of Section \ref{sec:general}). In fact, even if there are multiple solutions to (\ref{equ:hard_rods_example_F_0}) these solutions will be physically different since they give rise to different $d\expval{\lambda}$.

\subsection{Determining $d\expval{\lambda}$}
So far we have discussed how we can find a solution to the impurity problem given the effective Hamiltonian at $x\to -\infty$, in particular on $d\expval{\lambda}$. In fact, we do not know $d\expval{\lambda}$ a priori, but instead we need to compute it in a self consistent fashion: We need to match the assumed $d\expval{\lambda}$ with the $d\expval{\lambda} = \tfrac{1}{2\pi}1\upd{dr}d\int\dd{\mu}\mu n(x\to-\infty,\mu)$ computed from the solution.

First note that any solution to the fixed point equation (\ref{equ:hard_rods_example_F_0}) can always be upgraded to a full solution by a shift of the rapidity variable: In fact, if we define $v = \lambda-d\expval{\lambda}$ and parametrize the initial condition in terms of $n\ind{v}(v) = n(d\expval{\lambda}+v)$ the parameter $d\expval{\lambda}$ completely drops out of the equation. Then if we take any solution $n\ind{v}(x,\lambda)$ of the new  equation, compute its $\expval{\lambda} = \int\dd{v} n\ind{v}(v)$ and reintroduce the rapidity $\lambda = d\expval{\lambda}+v$ we always find a solution $n(x,\lambda)$ stationary GHD equation. 

Now let us consider a scattering scenario where we know the distribtion of incoming particles as a function $\lambda$. At this point we find that the problem is ill-defined. In order to specify which particles are incoming we need to know the rapidity corresponding to zero velocity $d\expval{\lambda}$.

Fortunately, in case of only incoming particles from the left there is a situation where we can uniquely specify the initial data. For $d>0$ the trick is to observe that the more particles are reflected the smaller $d\expval{\lambda}$ becomes. Thus $d\expval{\lambda}$ is largest for full transmission. Therefore, if we consider a state $n(\lambda)$ which is identically zero for all momenta smaller than some cutoff $\lambda_0$ and this cutoff is larger than $\lambda_0 > d\expval{\lambda} = \tfrac{d\int\dd{\lambda}pn(\lambda)}{1+d\int\dd{\lambda}n(\lambda)}$ then no matter the amount of reflected particles all incoming particles will always have positive velocity. Note that for $d<0$ we can always choose $\lambda_0=0$ since $d\expval{\lambda} \leq 0$. 

In the present case of an impurity which does not depend on $x$ and incoming particles only from the left, one can derive the following compact expression for the zero velocity rapidity:
\begin{align}
    d\expval{\lambda} = d\qty[N\upd{L}(\infty)-N\upd{L}(\bar{W})].\label{equ:hard_rods_example_dp}
\end{align}

Equation (\ref{equ:hard_rods_example_dp}) together with (\ref{equ:hard_rods_example_F_0}) are a two-dimensional closed system of equations:

By explicit computation one can check that the determinant of the Jacobian of this system is always positive if $d>0$. We conclude that for $d>0$ there always exists a unique solution to the scattering problem. 

\subsection{Hysteresis for negative hard rods $d < 0$}
\label{sec:hard_rods_hysteris}
\begin{figure}[!h]
    \centering
    \includegraphics[width=\textwidth]{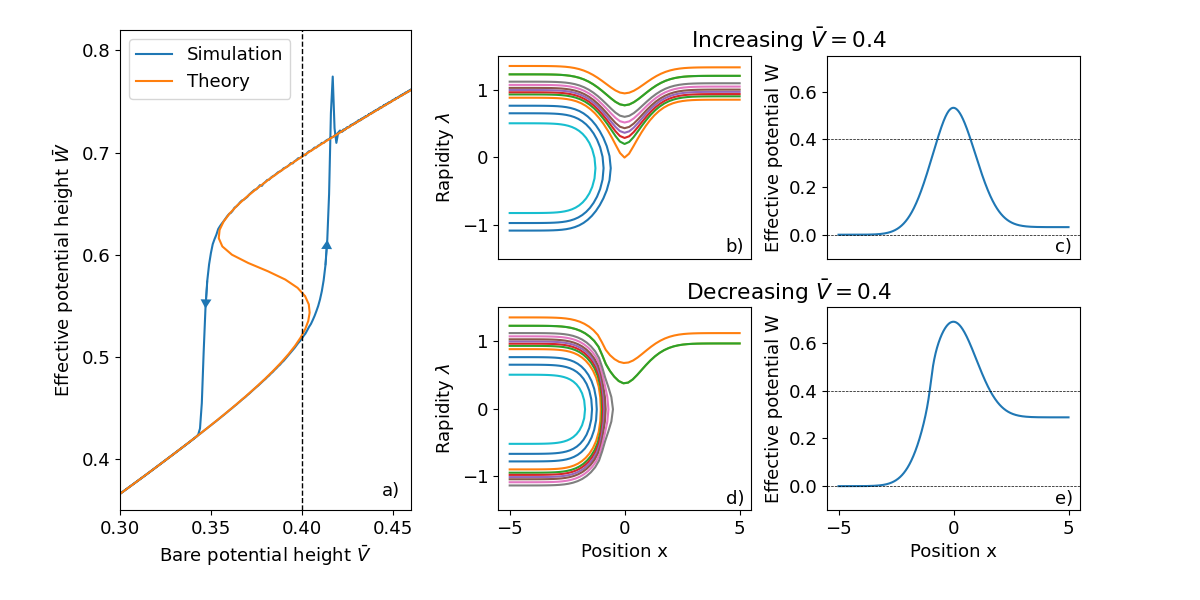}
    \caption{Hysteresis loop for $d=-0.3$ scattering of $n\ind{L}(\lambda) = \theta(\lambda-\tfrac{1}{2})\tfrac{1}{2}e^{-50(\lambda-1)^2}$ at a potential $V(x) = \bar{V}e^{-\tfrac{2}x^2}$: a) depicts part of a simulation, using the algorithm described in \ref{sec:simulation_direct}, where the bare potential height $\bar{V}$ is adiabatically ramped up and later down ($\Delta \bar{V} = 0.001$ per step). In the depicted range there are three theoretical solutions $\bar{W}$ (obtained by computing $\bar{V}$ as function of $\bar{W}$ numerically using (\ref{equ:hard_rods_example_dp}) and (\ref{equ:hard_rods_example_F_0})). The simulation follows the lower branch for increasing $\bar{V}$ and the upper branch for decreasing $\bar{V}$ (the middle branch is unstable). On the right side we picked $\bar{V}=0.4$ and show b), d) trajectories of particles at the impurity (obtained from the simulations) and c), e) the corresponding effective potential $W(x)$ as function of $x$ (computed from the simulation data via  \eqref{equ:hard_rods_position_impurity_dW_dx}).}
    \label{fig:hysteresis}
\end{figure}

Contrary to hard rods with positive size in the case of negative rod length there can indeed exist multiple solutions to the two-dimensional system of equations. We would like to demonstrate that using an explicit example. We study the scattering of $n\ind{L}(\lambda) = \theta(\lambda-\tfrac{1}{2})\tfrac{1}{2}e^{-50(\lambda-1)^2}$ at the potential $V(x) = \bar{V}e^{-\tfrac{x^2}{2}}$. We plot the values $\bar{W}$ of all solutions as function of $\bar{V}$ in Figure \ref{fig:hysteresis} (left) and compare it to a numerical simulation. In the range $\bar{V} \approx [0.36,0.41]$ there are three different solutions, only the outer two are stable. The numerical simulation was done using the algorithm described in \ref{sec:simulation_direct}. In addition we adiabatically ramp $\bar{V}$ up and later down: after each iteration we increase/decrease $\bar{V}$ ($\Delta \bar{V} = 0.001$ per step). For increasing $\bar{V}$ the simulation follows the lower branch, while for decreasing $\bar{V}$ the simulation follows the upper branch. On the right side of Figure \ref{fig:hysteresis} we also show the trajectories of particles and the shape of the effective potential. For small $\bar{V}$ only few particles are reflected, but as $1\upd{dr}(x) = \tfrac{1}{1+d\int\dd{\mu}n(x,\mu)} > 1$ the effective potential is already larger than the bare one by (\ref{equ:hard_rods_position_impurity_dW_dx}). When we increase $\bar{V}$ over the threshold value $\bar{V} \approx 4.1$ the situation suddenly changes: Now a considerable fraction of the particles are reflected, which increases $1\upd{dr}(x)$ on the left side, which in turn increases $W(x)$. Therefore even fewer particles can penetrate the impurity. This feedback loop only stops when most of the particles are reflected and only a negligible fraction of the particles is transmitted. If we now decrease $\bar{V}$ this situation is stable: The reflected particles contribute to a large $1\upd{dr}(x)$ and thus to a high potential $W(x)$, which most particles cannot penetrate. Only when we decrease $\bar{V}$ below the threshold value $\bar{V} \approx 3.6$ too many particles are transmitted and the system jumps back to the original configuration with few reflected particles. This explains the hysteresis observed in the numerical simulation.

\FloatBarrier

\section{Simulation of the GHD equation with impurity boundary condition}
\label{sec:comparison}

We would like to finish this paper with a demonstration that the GHD equation including the boundary conditions obtained from the mesoscopic impurity indeed gives the correct prediction for the evolution of the quasi-particle density in the large-scale limit $L \to \infty$.

We study the evolution of hard rods starting from an initial state characterized by (in macroscopic coordinates)
\begin{align}
    n(0,x\ind{macro},\lambda) = \theta(-3<x\ind{macro}<-0.5)\theta(0.5<\lambda<2) 7.5 e^{-\tfrac{1}{2}(x\ind{macro}+1.5)^2-\tfrac{25}{2}(\lambda-1)^2},\label{equ:hard_rods_comparision_init_state}
\end{align}
both via a simulation of the GHD equation and via direct simulation of the microscopic hard rods model. The impurity in this model is given by a potential $V(x\ind{micro}) = \alpha e^{-\tfrac{(x\ind{micro}/L\ind{imp})^2}{2}}$ (in microscopic coordinates), where $L\ind{imp} = \tfrac{1}{10}L^{\tfrac{3}{4}}$ is scaled with the macroscopic scale $L$. The relation between the microscopic and macroscopic coordinates is as follows: $x\ind{micro} = L x\ind{macro}$ and $t\ind{micro} = L t\ind{macro}$. We choose the height of the impurity to be $\alpha = 0.4$, the hard rod length to be $d=0.3$ and the simulation time to be $T\ind{macro}=3$, after which most of the particles have scattered.

The GHD simulation with impurity boundary condition are performed using the hybrid algorithm described in detail in \ref{sec:simulation_GHD_impurity}. Space, rapidity and time are discretized with $\Delta x\ind{macro} = 0.006, \Delta \lambda = 0.004$ and $\Delta t\ind{macro} = 0.001$ (for further details refer to \ref{sec:simulation_simulation}). At these values the GHD simulation has well converged. 

To simulate the microscopic hard rods simulation with an external potential we use a standard molecular dynamics algorithm, see for instance~\cite{biagetti2023threestage} (for simulating hard rods without external potential there is a more efficient algorithm~\cite{Doyon_2017,10.21468/SciPostPhysCore.3.2.016,10.21468/SciPostPhys.15.4.136,PhysRevLett.131.027101}). For a given $L$ we first distribute the hard rods randomly according to their initial distribution $\rho(0,x\ind{macro},\lambda) = \tfrac{1}{2\pi}1\upd{dr}(x\ind{macro})n(0,x\ind{macro},\lambda)$ and then evolve them time step by time step with $\Delta t\ind{micro} = 0.01$: The particles are simulated like independent particles using the Euler method, i.e. for the $i$'th particle at step $n$ we do $x_{i;n+1} = x_{i;n} + \lambda_{i;n} \Delta t\ind{micro}, \lambda_{i;n+1} = \lambda_{i;n} - \partial_x V(x_{i;n}) \Delta t\ind{micro}$. Whenever the distance between two particles becomes smaller than the rod size $d$, we displace both their trajectories by $d$, i.e.\ $x_{a,n} \to x_{a,n} + d$ and $x_{b,n} \to x_{b,n} - d$ for the scattering of particle $a$ with $b$ and $\lambda_a > \lambda_b$. Since we place the particles initially random, the resulting final state will also be random. Therefore, in order to gain sufficient statistics, this simulation is repeated $100$ times for each $L$ and averaged.

The results of both simulations are compared in Figure \ref{fig:mc_GHD_comparision}, where we plot a) the resulting $\rho(t\ind{macro}=3,x\ind{macro},\lambda)$ for both simulations as a density plot and b) the total density of particles in two regions R (reflected particles) and T (transmitted particles. Already in the density plot a) one can see that the quasi-particle density obtained via GHD with impurity boundary condition agrees quite well with the averaged density from the hard rods simulation. 

The results can also be compared more quantitatively in b). The data points and their errorbars are the mean and standard deviation of the hard rods Monte Carlo results for several $L$. We can see that this agrees well with the GHD simulation (solid line) in both regions. 

We conclude that the GHD  equation with impurity boundary conditions indeed correctly describes the large scale evolution of an integrable model with a mesoscopic impurity.

Further numerical results are given in \ref{sec:simulation_simulation}: We simulate the hard rods GHD equation starting from the same initial state and the same impurity, but for different hard rod lengths $d=-0.3,0, 0.3$ and qualitatively discuss the difference of how they scatter at the impurity.

\FloatBarrier

\section{Conclusion}
In this paper we studied mesoscopic impurities in the GHD framework. First, we discussed how to include impurities into GHD in general and found that they are described by boundary conditions. The boundary conditions correspond to non-equilibrium stationary states of the microscopic impurity model. A big complication in interacting models compared to non-interacting models is that reflected particles affect the incoming particles (in particular their effective velocities) and therefore one cannot specify the incoming particles without knowing the outgoing particles. This means the solution to the scattering problem is only given by a collection of possible solutions, but it is not directly given as a map from the incoming data to the outgoing data.

Since it is not known how to find non-equilibrium stationary states in general, we decided to study a specific class of impurities: Impurities on a mesoscopic scale (larger than the diffusive scale), which are linear combinations of charge densities. These can be described by the (Euler-scale) stationary GHD equation in an external potential. This provides a broad class of impurities, which are present in all integrable models and can be studied analytically also for strong impurities.

Additional simplification occurs when we restrict to models where the scattering phase shift is only a function of the difference of the momenta. Here the scattering problem can be interpreted as particles moving in a one-particle effective Hamiltonian $H(x,\lambda)$, given by the bare energy $E(x,\lambda)$ plus an interaction dependent correction. Given the incoming data we described how to solve the scattering problem at such a Hamiltonian $H(x,\lambda)$ in general. This allows us to write down a functional fixed point equation for the correction term in $H(x,\lambda)$. If the incoming data is known then a solution to the fixed point equation corresponds to a valid solution to the stationary GHD equation. However, as the outgoing data might affect the incoming data, in general a solution to the fixed point equation will not reduce to the assumed incoming state for $x \to \pm \infty$. Instead, one would have to adapt the asymptotic data until it is matched by the incoming data of the solution to the fixed point problem.

Still, the fixed point equation provides a starting point to study mesoscopic impurities in general and even to find analytic solutions to the scattering problem. In general, mesoscopic impurities show the following features: They are invariand under local spatial rescaling in $x$, the scattering at them is deterministic and scattering vanishes for sufficiently weak impurities. Furthermore, it is possible to solve the scattering problem for mesoscopic impurities via an efficient algorithm.

We provided almost explicit solutions to an impurity in the hard rods model, given by a potential, which is independent of $\lambda$. Here the difference between a positive (corresponding to a time delay) and a negative (corresponding to an instant jump) scattering phase shift $\varphi(\lambda)$ becomes apparent: Negative $\varphi(\lambda)$ tends to increase the transmission, while positive $\varphi(\lambda)$ decreases it. For the hard rods we also compared molecular simulations and GHD simulations with impurity and found agreement. This shows that the scattering can be indeed captured by a boundary condition.

Mesoscopic impurities are of course only an approximation for large impurities, in particular since we only work on the Euler scale. It would be interesting to go beyond that. In regard of the gradient expansion we can view this as a first order approximation in the impurity size $1/L\ind{imp}$. By adding more terms of the gradient expansion one can compute higher order corrections to our results which would allow the treatment of smaller mesoscopic impurities (for instance on the diffusive scale). However, it is not clear whether this expansion extends all the way to microscopic impurities, i.e.\ whether this expansion converges or is only an asymptotic expansion. To study this we expect that it will be interesting to look at the transition between deterministic and non-deterministic scattering: We have established in this paper that scattering at (Euler scale) mesoscopic impurities is always deterministic: particles are always transmitted or reflected with probability one. This is of course not generally true for microscopic impurities. It would be interesting to see whether higher order gradient expansion~\cite{10.21468/SciPostPhys.6.4.049,PhysRevLett.121.160603,DeNardis_2023} introduces non-deterministic scattering, otherwise microscopic impurities will be fundamentally different from mesoscopic ones. 

In general it would be interesting to also gain a deeper understanding of microscopic impurities in interacting models and in particular the structure of the resulting GHD boundary conditions. We have already observed for mesoscopic impurities that the solution to the scattering problem is not unique. If this case occurs during a (GHD + impurity boundary condition) simulation one would need to find a way to choose the correct impurity solution out of multiple ones. In the situation we discussed in Section \ref{sec:hard_rods_hysteris} the correct solution was identified by following the solution set adiabatically. Also, the numerical algorithm outlined in \ref{sec:simulation} is designed to pick the physical solution. For microscopic impurities there could potentially be plenty (or even a continuum) of solutions, meaning that it might be impossible to identify the physical one. In this case the GHD equation + boundary condition does not determine the large scale dynamics completely. To resolve this problem one would either have to compute further information about the state at the impurity, or alternatively try to find a physical principle on how to identify the correct solution. In our opinion one candidate for this is the principle of maximum entropy: From an old-school naive thermodynamic viewpoint, in case of multiple possible solutions the physical solution should be the one that maximizes entropy. This kind of reasoning has been, for instance, used to identify the physical weak solution after a shock in hyperbolic conservation laws (the stereotypical example of this is Burgers' equation)~\cite{Bressan2013}. It is not clear to us, however, how such a principle could be applied to impurity boundary conditions. The complication is that the impurity is not a closed system: During scattering entropy is constantly transported into and out of the impurity. A principle of entropy maximization would require to take into account these non-vanishing entropy currents. We believe that is an interesting direction for further research.

But also on the level of mesoscopic impurities there are open questions left. We discussed that solutions to the scattering problem are not unique in general, but it is not clear whether this appears in all models and under which conditions. In particular, it would be interesting to have a more detailed look at the scattering problem in quantum mechanical models, for instance the Lieb-Liniger model. There the interaction between particles is rapidity dependent and also $\varphi(\lambda)$ is integrable which is quite different from the hard rods model. Furthermore, we restricted to impurities which can be described by external potentials. Another possibility would be to have an impurity where the interaction changes locally. If the interaction changes sufficiently slowly, one can describe the system by GHD~\cite{PhysRevLett.123.130602} and therefore it should be possible to study properties of such impurities using similar ideas as for external potentials.

It would also be interesting to go beyond these simple integrable models and also consider models where $\varphi(\lambda,\mu) \neq \varphi(\lambda-\mu)$ is not a function of the difference of momenta only. In this case, the normal modes of the GHD equation with external potential are not known, so most of the analysis in this paper cannot be applied. It is not even clear whether  normal modes will always exist in any model. If there are models where they do not exist, scattering could be quite different. For instance, since one cannot think about the system as particles moving along GHD characteristics anymore, scattering might be non-deterministic. Also the numerical algorithms outlined in \ref{sec:simulation} do not apply in that case and it would be interesting to extend them.

A completely different direction of research would be to implement mesoscopic impurities in actual experiments. We believe that this should be possible among others in cold atom experiments, which are well described by the Lieb-Liniger model. For instance, a potential barrier could be implemented similar to an external potential, which is already used in GHD experiments~\cite{Kinoshita2006,PhysRevLett.122.090601}. A precise setup we have in mind would be to prepare the atoms in a 1D trap (similar to~\cite{PhysRevLett.122.090601}). Then at time $t=0$ the trap is removed and at the same time a sharply peaked potential is turned on which serves as the impurity. After releasing the atoms, the distribution of particles can be observed as function of time and then compared to theoretical computations. A feature that would be particularly interesting for experimental realization is the local rescaling property of mesoscopic impurities. This implies some kind of stability against perturbations: Only few specifications of the impurity determine the scattering behavior, the precise shape of the impurity is not important.

\ack
I would like to thank Benjamin Doyon for reading the manuscript, his helpful comments and discussing the topic with me. Funding from the faculty of Natural, Mathematical \& Engineering Sciences at King's College London is acknowledged. Numerical simulations were done using the CREATE cluster~\cite{CREATE}.

\appendix
\setcounter{section}{0}

\section{Numerical simulation}
\label{sec:simulation}
\subsection{Scattering problem of GHD}
The aim of this appendix is to establish an efficient algorithm which allows to solve the scattering problem at a potential in GHD at a given potential. By that we mean solving
\begin{align}
    (\partial_\lambda E)\upd{dr}\partial_x n &= (\partial_x E)\upd{dr}\partial_\lambda n,\label{equ:numerics_scattering_GHD}
\end{align}
given some incoming state on the left and on the right. Again as we discussed in Section \ref{sec:general} one cannot specify which particles are incoming, as it depends on the reflected/transmitted particles. However, for numerical purposes this is not too relevant as particles with the wrong velocity will simply move away from the potential and thus will not affect the scattering at the potential.

\subsubsection{A naive algorithm}
Standard algorithms to solve equations of motion are based on finite time step schemes: One takes the initial state and evolves it by a small time $\Delta t$. This gives another state and by iterating this procedure $T/\Delta t$ times one finally reaches time $T$. These algorithms usually are exact as $\Delta t \to 0$ and thus this procedure allows to find a good numerical approximation to the actual solution if $\Delta t$ is small enough. For a review of higher order numerical methods for the GHD equation see~\cite{møller2022dissipative}.

For the scattering problem this is not possible simply because time does not appear in (\ref{equ:numerics_scattering_GHD}). In case there is no reflection and only incoming particles from the left, one could use $x$ as `time'-variable. By that we mean start at $x\ind{init} \ll 0$ and then propagate $n(x,\lambda)$ from $x$ to $x+\Delta x$ until one reaches a large $x\ind{final} \gg 0$. Unfortunately, in case there is reflection we do not know how much is coming back, thus an algorithm of that kind is not possible.

What is possible, however, is to simulate the time-dependent GHD equation:  Recall the derivation of (\ref{equ:numerics_scattering_GHD}), see Section \ref{sec:mesoscopic}, and reinstate time as in equation (\ref{equ:mesoscopic_derivation_GHD_rescaled}). One interpretation of equation (\ref{equ:mesoscopic_derivation_GHD_rescaled}) is that the limit $\ell \to \infty$ corresponds to a long time limit. Thus, if we simulate the ordinary GHD equation for a long time, the state will approach a stationary state, which is a solution to the stationary GHD equation. Therefore, a naive algorithm to numerically obtain the stationary state is to use a finite time step scheme to compute the solution at a long time $T$ and then send $T \to \infty$ until convergence is reached. 

An algorithm of this kind of course theoretically works, but in practice it is very inefficient. The problem is that due to the long time $T$, errors from each step will add up. This means in order to keep precision one has to choose $\Delta t$ smaller every time one increases $T$, which leads to an even larger number of steps. Furthermore let us note that simulating a single step of the GHD equation is relatively time consuming since one has to compute the dressing of $\partial_\lambda E$ and $\partial_x E$, which involves solving an infinite-dimensional (or very high dimensional after discretization) linear equation at each point in space.

\subsubsection{Direct simulation of the stationary GHD equation}
\label{sec:simulation_direct}
The above algorithm is inefficient because it does not make any use of the special properties of the stationary state. Now we will describe a more efficient algorithm which is based on the idea that the stationary state can be described by particles evolving according to some Hamiltonian, defined via (\ref{equ:stationary_effective_hamiltonian_GHD}). Given $(\partial_\lambda E)\upd{dr}$ and $(\partial_x E)\upd{dr}$ the solution $n(x,\lambda)$ can easily be found by computing the characteristics (\ref{equ:mesoscopic_stationary_ode}). Given the characteristics we can compute $(\partial_\lambda E)\upd{dr}$ and $(\partial_x E)\upd{dr}$. Therefore, an iterative algorithm to solve the stationary GHD equation given the distribution of incoming particles is as follows:
\begin{enumerate}
    \item Guess an initial $n(x,\lambda)$, for instance $n(x,\lambda) = 0$
    \item Use $n(x,\lambda)$ to compute $(\partial_\lambda E)\upd{dr}$ and $(\partial_x E)\upd{dr}$
    \item Choose an $x_0$ s.t. $x_0$ is outside the impurity. For a set of initial momenta $\qty{\lambda_0}$ compute the characteristics starting at $x(0)=\pm x_0$ and $\lambda(0)=\lambda_0$.
    \item From the characteristics compute a new $n(x,\lambda)$ by transporting the incoming data along the characteristics
    \begin{align}
        n(x(t),\lambda(t)) = n(x(0),\lambda(0)).
    \end{align}
    Here $n(x(0),\lambda(0))$ is the incoming data.
    \item Repeat from step 2 using the new $n(x,\lambda)$. Iterate this procedure until convergence.
\end{enumerate}

The above algorithm is quite efficient: Each iteration still consists of two time consuming steps, computing the characteristics and computing $(\partial_\lambda E)\upd{dr}$ and $(\partial_x E)\upd{dr}$. However, in practice we find that the algorithm usually converges relatively fast (within $\sim 10$ iterations) and thus this is a huge speed up compared to the simulation of the time-dependent GHD simulation. Another advantage of this algorithm is that it gives access to the characteristics and in particular directly shows which quasi-particles are reflected and which are transmitted. Also from $(\partial_\lambda E)\upd{dr}$ and $(\partial_x E)\upd{dr}$ one can reconstruct the effective Hamiltonian (\ref{equ:stationary_effective_hamiltonian_GHD}).

Remark: The precision of the algorithm is of course greatly influenced by the choice of the set of initial momenta $\qty{\lambda_0}$ for the characteristics. Note that not only the total number of characteristics are important, but also their distribution. In fact, it is important to make sure that there are sufficient characteristics in regions where the incoming densities $n(\pm x_0,\lambda_0)$ are high. High densities have the biggest impact on the precision of the dressing, which in turn is important to be able to compute accurate characteristics. On the other hand it is also important to have at least some characteristics in low density regions. We found that the following way of distributing produces good results: First, we sample the majority of the $\lambda_0$ from the probability measures which are proportional to the incoming densities. This naturally places many characteristics in high density regions, but places only few characteristics in the low density regions. In order to ensure sufficient coverage of those regions as well we place the remaining characteristics with a uniform spacing between them, i.e.\ $\lambda_0 = k \Delta \lambda$, where $k \in \mathbb{N}$.

In our simulations we choose a small time-step $\Delta t = 0.001$ to compute the characteristics. The dressed $(\partial_\lambda E)\upd{dr}$ and $(\partial_x E)\upd{dr}$ we compute by partitioning the space into small boxes of size $\Delta x = 10 \Delta t$. Therefore, each characteristic will provide $\sim 10$ points $(x(t),\lambda(t))$ in each box. We collect all those points from all characteristics and sort them according to their momenta. At these points $(x(t),\lambda(t))$ we now know the value of $n(x(t),\lambda(t))$. In order to find an approximation for the full $n(x,\lambda)$ we (linearly) interpolate between those base points.

\subsection{Simulating the GHD equation with mesoscopic impurity}
\label{sec:simulation_GHD_impurity}
So far we studied how to solve the scattering problem in GHD. As we discussed this produces the boundary conditions for the GHD equation. In this section we now want to include these boundary conditions into a simulation of the GHD equation. That is, given some initial macroscopic distribution $n_0(x,\lambda)$, how does $n(t,x,\lambda)$ evolve in time in the presence of the impurity. The equation we want to solve is given by:
\begin{align}
    \partial_t n &= -v\upd{eff} \partial_x n & x&\neq 0,\label{equ:numerics_wavepacket_free_GHD}
\end{align}
and the corresponding boundary conditions at $x=0$. Note that in a situation without impurity there are methods to obtain the solution directly at time $t$~\cite{DOYON2018570}. Unfortunately, this is not possible with impurity since the outgoing particles of the impurity have to be taken into account at each time step and this information cannot be directly obtained from the initial data. Instead one has to proceed time step by time step.

Outside the impurity one can simulate the GHD equation using ones preferred algorithm (we simply discretize space, momenta and time to first order and evolve using the right hand side of the GHD equation (\ref{equ:numerics_wavepacket_free_GHD})). In between each step one has to solve the impurity problem using the current incoming data on the left and on the right side of the impurity. There are two options: One way is to make use of the fixed point problem we introduced in this paper and analytically solve for the solution. If possible this method can be very fast. For instance, for the hard rods with $d>0$ we already showed that there always exists a unique solution. But in a general model the analytical solution of the scattering problem becomes way more complicated. Also it is not clear whether the solution will be unique and thus, in case of multiple solutions, it is not clear which solution to choose.

Instead, one can solve the impurity problem using the iterative scheme described in the last section. This algorithm has the advantage that it can be implemented for any model. In addition, using the following simple trick the algorithm will automatically deal with the possibility of multiple solutions: Instead of solving the impurity problem from scratch during each time step, one can start the iterative procedure from the solution obtained in the last time step. This way the state at the impurity will adiabatically follow the correct solution to the impurity problem over time. Furthermore, we find that it is sufficient to perform one iteration of the iteration scheme in each time-step. This gives an efficient hybrid algorithm which solves the GHD equation with impurity. For now this algorithm is first order, but it would be interesting to extend it to higher order methods as well~\cite{møller2022dissipative}.

\subsection{Comparison of the hard rods scattering for different $d$}
\label{sec:simulation_simulation}

We now use the hybrid algorithm described in the last section to simulate an explicit example: hard rods scattering at a mesoscopic potential $V(x) = \alpha e^{-\tfrac{1}{2}x^2}$ for different rod lengths $d=-0.3,0,0.3$. For $d=0.3$ this is the same simulation as in Section \ref{sec:comparison}. For the part outside the impurity we use a basic algorithm: We discretize space and rapidity on a $1000 \times 1000$ grid with spacings $\Delta x\ind{macro} =  0.006$, $\Delta \lambda =  0.004$ and store $n_{ij} = n(x_i,\lambda_j)$. We also discretize time with $\Delta t\ind{macro} = 0.001$. In each step we numerically compute the effective velocity $v\upd{eff}(x,\lambda)$ using (\ref{equ:hard_rods_eff_velo}) and the derivative $\partial_x n$ and update $n(t+\Delta,x,\lambda) = n(t,x,\lambda) - v\upd{eff}(t,x,\lambda)\partial_xn(t,x,\lambda)$, except for the impurity site. At the impurity we simulate the trajectories of $N = 5000$ hypothetical particles with initial momenta equally spaced between $\lambda = 0.4$ and $\lambda=2$ by doing one iteration of the algorithm described in \ref{sec:simulation_direct}. Initially the impurity is empty. After each step we store the computed $\lambda\upd{dr}$ and $1\upd{dr}$ and use it for the simulation of particle trajectories in the next step. The endpoints of the particle trajectories give a distribution of outgoing particles, which we feed back into the simulation outside of the impurity.

\begin{sidewaysfigure*}
    \centering
    \vspace{-16cm}
    \includegraphics[width=\textwidth]{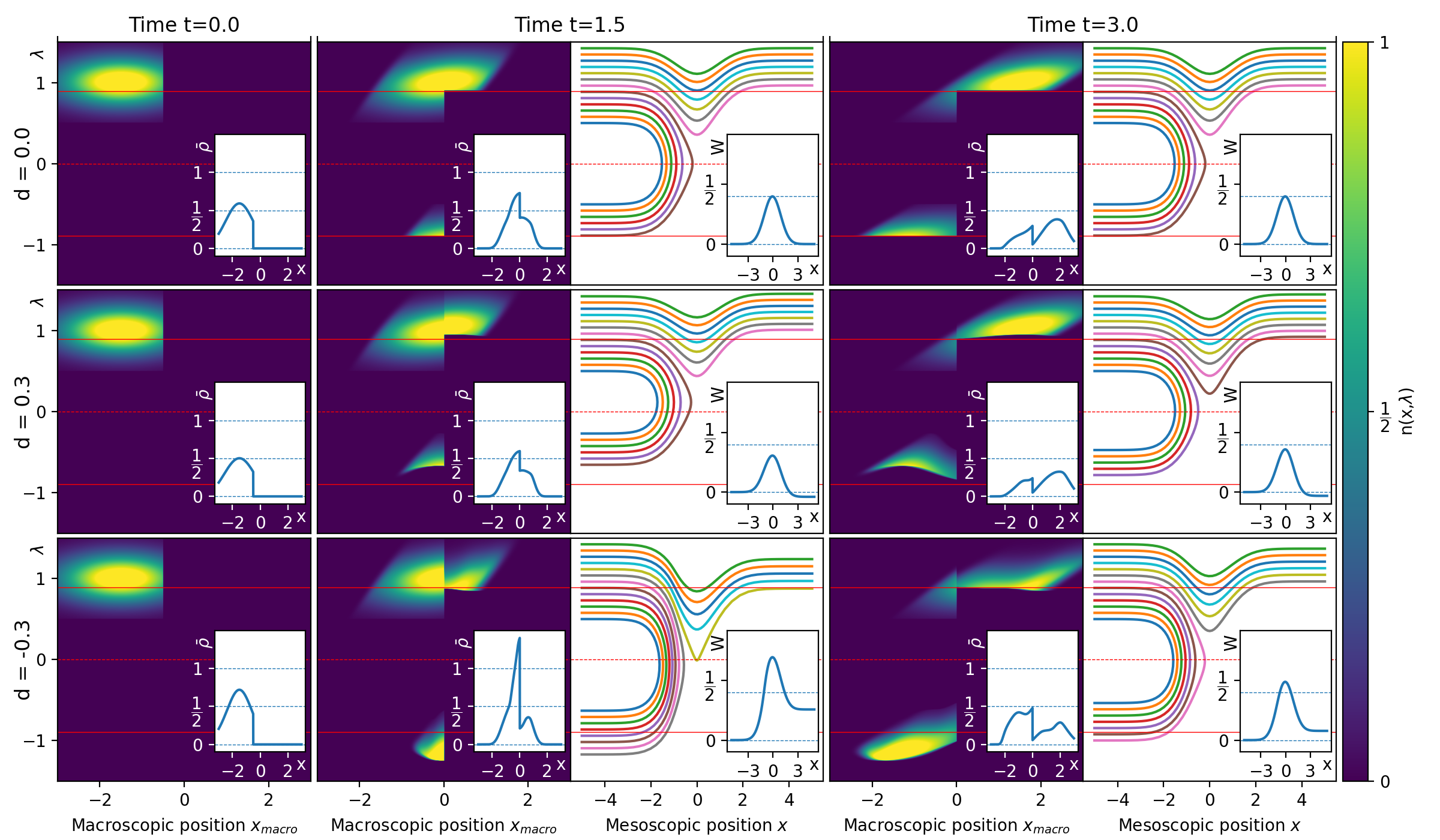}
    \caption{Simulation of hard rods scattering at an impurity $V(x) = \alpha e^{-\tfrac{1}{2}x^2}$ with $\alpha = 0.4$, for $d=0$ (non-interacting), $d=0.3$ and $d=-0.3$. The density plots show the distribution of $n(t,x\ind{macro},\lambda)$ for $t=0$, $t=1.5$ and $t=3$ and the inset depict the rapidity integrated density $\bar{\rho}(x\ind{macro})$. For time $t=1.5$ and $t=3$ we also plot some trajectories of particles at the impurity, together with the current effective potential $W(x)$ (inset). The red lines at $\lambda_0 = \pm\sqrt{2\alpha}$ (solid) and $\lambda=0$ (dashed) are given for comparison with the non-interacting case.}
    \label{fig:simulation_wavepacket}
\end{sidewaysfigure*}

In Figure \ref{fig:simulation_wavepacket} we give the result of the simulation for $d=0$ (non-interacting particles), $d=0.3$ and $d=-0.3$. The height of the potential is $\alpha = 0.4$. The initial state at $t=0$ is given by (\ref{equ:hard_rods_comparision_init_state}). We show the density plots at 3 different times, $t = 0$, $t=1.5$ and $t=3$. In addition for $t=1.5$, and $t=3$ we also depict the trajectories at the impurity and the effective potential $W(x)$ (inset). The red solid line indicate the momenta $\lambda_0 = \sqrt{2\alpha}$ whose kinetic energy equal the height of the potential in the non-interacting case. For non-interacting particles, particles with a higher momenta are transmitted and particles with lower momenta are reflected. The scattering of interacting particles is different:

For positive $d=0.3$, since $1\upd{dr}(x) < 1$, the effective potential $W(x)$ is smaller than the bare potential $V(x)$ (see equation \ref{equ:hard_rods_position_impurity_dW_dx}). Thus, more particles can penetrate the impurity. Also since $W(\infty)<0$ the transmitted particles have higher momenta than the corresponding incoming ones.

For negative $d<0$ we have $1\upd{dr}(x) > 1$ and thus the effective potential is increased. In the beginning of the simulation when the density of particles is very low, particles are transmitted. At time $t=1.5$ the density at the impurity is so high that most particles are reflected. Only towards the end of the simulation, at $t=3$, the density decreases again, which decreases the effective potential and allows more particles to be transmitted.

\section{Explicit construction of a solution for weak impurities}
\label{sec:weak_proof}
In this appendix we give the explicit construction of the solution for a sufficiently weak impurity discussed in Section \ref{sec:weak}, i.e.\ we want to show the following observation:
\obsweak
The computations below are a bit technical, the idea, however, is very simple. Since the impurity is weak the level sets of the effective Hamiltonian (i.e.\ the trajectories of particles) will be almost straight. Therefore no reflection occurs. This allows for the following construction:

Let us call $\lambda\ind{L}(-\infty)$ the smallest incoming rapidity from the left and $\lambda\ind{R}(\infty)$ the smallest incoming rapidity from the right. Then there will be two curves $\lambda\ind{L}(x)$ and $\lambda\ind{R}(x)$ which separate the space into three distinct parts: In the region $\lambda>\lambda\ind{L}(x)$ we have particles coming from the left to the right and in the region $\lambda<\lambda\ind{R}(x)$ we have particles coming from the right. In the region $\lambda\ind{R}(x)<\lambda<\lambda\ind{L}(x)$ the density vanishes. Note that this construction only works if there are no reflected particles. 

For convenience let us redefine the Hamiltonian by adding a constant, s.t. $H(-\infty, v_\lambda=0) = 0$. Also let us redefine:
\begin{align}
    N\upd{L}(h) &= \int_{H(-\infty,\lambda\ind{L}(-\infty))}^h\dd{h} n(-\infty, H^{-1}(-\infty,h))\\
    N\upd{R}(h) &= \int_{H(-\infty,\lambda\ind{R}(-\infty))}^h\dd{h} n(-\infty, H^{-1}(-\infty,h)),
\end{align}
where in both definitions we use different branches of $H^{-1}(-\infty,\cdot)$. Then we can write:
\begin{align}
    N(x,\lambda) = \begin{cases}
        N\upd{L}(H(x,\lambda)) &\lambda>\lambda\ind{L}(x)\\
        N\upd{R}(H(x,\lambda)) &\lambda<\lambda\ind{R}(x)\\
        0 & \lambda\ind{R}(x)<\lambda<\lambda\ind{L}(x).
    \end{cases}
\end{align}

We now have to find a solution to the adapted fixed point equation:
\begin{align}
    U(x,\lambda) = \Big[&\int_{\mu>\lambda\ind{L}(x)}\dd{\mu}T(\lambda-\mu) N\upd{L}(H(-\infty,\mu) + \alpha V(x,\mu) + U(x,\mu))\nonumber\\
    &+\int_{\mu<\lambda\ind{R}(x)}\dd{\mu}T(\lambda-\mu) N\upd{R}(H(-\infty,\mu) + V(x,\mu) + U(x,\mu))\nonumber\\
    &- \int_{\mu>\lambda\ind{L}(-\infty)}\dd{\mu} T(\lambda-\mu) N\upd{L}(H(-\infty,\mu))\nonumber\\
    &- \int_{\mu<\lambda\ind{R}(-\infty)}\dd{\mu} T(\lambda-\mu) N\upd{R}(H(-\infty,\mu))\Big].\label{equ:stationary_small_fixed_point}
\end{align}

Note that $\lambda\ind{L/R}$ implicitly depend on $U(x,\lambda)$. For $\alpha = 0$ a solution to this equation is simply given by $U(x,\lambda) = 0$ and $\lambda\ind{L/R}(x) = \lambda\ind{L/R}(-\infty)$. We already established in Section \ref{sec:stationary_uniqueness} that the Jacobian is given by the inverse dressing operation. Because the dressing has to be defined for $\alpha = 0$, we know that the Jacobian is invertible at $\alpha = 0$. Therefore, there will be a small region $|\alpha| < \alpha_1$ in which a solution to the fixed point equation will exist (mathematically this follows from the implicit function theorem), which will also be close to the unperturbed solution $|U|\leq U_0 = \order{\alpha}$. This establishes, for sufficiently small $\alpha$, the existence of the effective Hamiltonian $H(x,\lambda)$ describing the scattering situation (even though we do not know its explicit form).

However, this Hamiltonian was found on the assumption that there is no reflection. For consistency we now need to check that this Hamiltonian $H(x,\lambda) = H(-\infty,\lambda) + \alpha V(x,\lambda) + U(x,\lambda)$ does not have any critical points in the regions $\lambda>\lambda\ind{L}(x)$ and $\lambda<\lambda\ind{R}(x)$ (the existence of such a critical point is necessary for reflection). Without loss of generality consider the region $\lambda>\lambda\ind{L}(x)$ (the other one can be treated equivalently). The boundary of the region is determined via:
\begin{align}
    H(x,\lambda\ind{L}(\infty))= E_0(\lambda\ind{L}(\infty)) + \alpha V(x,\lambda\ind{L}(\infty)) + U(x,\lambda\ind{L}(\infty)) = E_0(\lambda\ind{L}(-\infty)),
\end{align} 
since $\lambda\ind{L}(x)$ follows a level set of $H$. By definition any critical point has to satisfy $\partial_\lambda H(x,\lambda) = 0$:
\begin{align}
    \partial_\lambda E_0 + \alpha \partial_\lambda V(x,\lambda) + \partial_\lambda U(x,\lambda) = 0.
\end{align}

In case $\alpha = 0$ then $\lambda\ind{L}(x) = \lambda\ind{L}(-\infty)$ and $\partial_\lambda E_0(\lambda\ind{L}(-\infty)) \geq C$. By continuity, if $\alpha$ is small (and therefore $U$ is small) there will be a region $|\alpha| < \alpha_2$ in which $|\alpha \partial_\lambda V(x,\lambda) + \partial_\lambda U(x,\lambda)| < C$ and therefore $H(x,\lambda)$ does not have any critical points for $|\alpha| < \alpha_2$.

The same is true for $\lambda<\lambda\ind{R}(x)$ and thus our ansatz of $H(x,\lambda)$ is consistent: For sufficiently small $\alpha$ there cannot be any critical points and therefore no `islands' $\Lambda\upd{L/R}_i$ and also no reflection. 

This finishes the construction of the solution to the scattering problem for sufficiently small $\alpha$. We now go on to show that for this solution the state on the left and on the right side are actually coincide (meaning that all particles just flow `around' the impurity, without being affected by it). Note that this is not trivial: For instance, one could imagine that all particles are transmitted but their final momenta might be shifted compared to their initial momenta. In order to study this let us look again at equation (\ref{equ:stationary_small_fixed_point}). This fixed point equation is in fact independent for different $x$, which -- in this case -- holds globally since there are no critical points of the Hamiltonian in the regions where particles are. Now imagine starting at $x = -\infty$ and observing how the fixed point $U(x,\lambda)$ varies as $x$ runs from $x =-\infty$ to $x =\infty$. $U(x,\lambda)$ change adiabatically as function of $\alpha V(x,\lambda)$, but will stay in some finite region $|U| \leq U_0$, where $U_0$ depends on $\alpha$. Again, since the Jacobian is given by inverse dressing operation, and the dressing is invertible at $V = 0$, the Jacobian will be invertible for all impurity potentials $V(x,\lambda)$ in a  neighbourhood around $V = 0$ as well (mathematically this follows from the inverse function theorem). In this neighbourhood the solution $U(x,\lambda)$ will be a unique function $U(x,\lambda) = U[\alpha V(x,\lambda)](x,\lambda)$ of $\alpha V(x,\lambda)$. If $\alpha$ is small enough $|\alpha| < \alpha_3$ then $\alpha V(x,\lambda)$ will stay inside this neighbourhood. Since for $x \to \infty$ we have that $V(x,\lambda) \to 0$, far away from the impurity the solution is given by $U(x \to \infty,\lambda) = U[0](x,\lambda)$. By construction however, we know that $U[0](x,\lambda) = 0$. Therefore, the Hamiltonians on the left and the right side will be equal, which in turn implies the same for the states on both sides. This concludes the derivation.

\section*{References}
\bibliography{refs.bib}

\end{document}